\documentclass[aj_pt4]{aastex}

\newcommand{\be}{\begin{equation}}
\newcommand{\ee}{\end{equation}}

\slugcomment{}

\shorttitle{Local Surface Density}
\shortauthors{V. I. Korchagin, T. M. Girard}

\begin{document}


\title{Local Surface Density of the Galactic Disk from a 3-D Stellar Velocity Sample  
       }
	

\author{V.I. Korchagin\altaffilmark{1,2}and T.M. Girard \altaffilmark{1}}
\affil{Yale University, Dept. of Astronomy, P.O. Box 208101, 
New Haven, CT 06520-8101
}
\email{vik@astro.yale.edu}
\email{girard@astro.yale.edu}

\author{T. V. Borkova }
\affil{Institute of Physics, Rostov-on-Don 344090,  Russia;
Isaac Newton Institute of Chile, Rostov-on-Don Branch}
\email{borkova@rsusu1.rnd.runnet.ru}

\author{D. I. Dinescu  \altaffilmark{3} and W. F. van Altena}
\affil{Yale University, Dept. of Astronomy, P.O. Box 208101, 
New Haven, CT 06520-8101
}
\email{dana@astro.yale.edu}
\email{vanalten@astro.yale.edu}

\altaffiltext{1} {Equal first author}

\altaffiltext{2}{also Institute of Physics, Rostov University, Rostov-on-Don 344090, Russia;
Isaac Newton Institute of Chile, Rostov-on-Don Branch}

\altaffiltext{3}{also Astronomical Institute of the Romanian Academy, Str. Cutitul de Argint 5, 
RO-75212, Bucharest 28, Romania}

\begin{abstract}

We have re-estimated the surface density of the Galactic disk in the solar neighborhood
within $\pm$ 0.4 kpc of the Sun 
using parallaxes and proper motions of a kinematically and spatially unbiased
sample of 1476 old bright red giant stars from the Hipparcos
catalog with measured radial velocities from Barbier-Brossat \& Figon (2000). 
We determine the vertical distribution of the red giants as well as the
vertical velocity dispersion of the sample, (14.4 $\pm$ 0.3 km/sec), and
combine these to derive the surface density of gravitating
matter in the Galactic disk as a function of the Galactic coordinate $z$.
The surface density of the disk increases from 10.5  $\pm$ 0.5 $M_{\odot}$ / pc$^2$ 
 within $\pm$ 50 pc
to 42 $\pm$ 6 $M_{\odot}$ / pc$^2$ within $\pm$ 350 pc. The estimated volume density
of the Galactic disk within $\pm$ 50 pc is about 0.1 $M_{\odot}$ / pc$^3$
which is close to the volume density estimates of the 
observed baryonic matter in the solar neighborhood.

\end{abstract}


\section{Introduction}

Starting with the pioneering works by Schwarzschild, Lindblad,  and Oort,
the study of local stellar kinematics is one of the priorities in astronomy.
Such studies provide important information about the structure and evolution
of the Milky Way. One of the fundamental characteristics 
 is the total density of gravitating mass of the Milky Way disk.
Knowledge of this value allows us to make a conclusion as to
the presence of dark matter in the Galactic disk 
by comparing it with the observed density of visible matter.
Oort (1932, 1960) first determined the total column density of mass in the
vicinity of the Sun 
based on the assumption that motion of stars perpendicular
to the Galactic plane can be separated from motions of stars in the plane.
He found the value  $\Sigma$ for $ |z| <$ 700 pc is approximately 90 M$_{\odot}$ /pc$^2$.
The total volume density near the Sun estimated by Oort (1932) is 0.092 M$_{\odot}$ /pc$^3$,
indicating the presence of a considerable amount of dark matter
as compared to the adopted volume density of stars in the solar neighborhood of
0.038 M$_{\odot}$ /pc$^3$.  The problem was re-analyzed later in several ways.
As noticed by Kuijken \& Gilmore (1989a,b), there are two  
different approaches to determine the mass density in the solar neighborhood. The first approach
determines the local volume mass density or Oort's limit, while the other 'experiment'
measures the surface density of the disk integrated over a range of vertical heights $z$.
The local volume density determinations are based on the assumption that
the vertical motions of stars in the disk are decoupled from their motions
in the plane (see, however, the discussion of this assumption by Statler (1989)).
The problem can then 
be reduced to the integration of the velocity distribution function
over the vertical velocity component of stars (Fuchs \& Wielen 1993, Flynn \& Fuchs 1994).
Main sequence are usually used in such an approach to be representative
of the total stellar component, and the result needs only to be corrected
for brighter main sequence stars and giants (Flynn \& Fuchs 1994).
The volume density estimate based on this method
 was used by Flynn and Fuchs (1994), Cr\'{e}z\'{e} et al. (1998)
and Holmberg and Flynn (2000). These authors found that the dynamical volume density
in the solar neighborhood is in agreement with the observed local mass density
estimates, and that there is no compelling evidence for significant amounts of
dark matter in the local disk.

Kuijken \& Gilmore in their series of papers (1989a, 1989b, 1991)
used a sample of K-dwarfs as a stellar tracer population to determine the surface density
of the Milky Way disk.  They found that the integral
surface mass density of all mass (disk + halo) within 1.1 kpc of the Sun is
71 $\pm$ 6 M$_{\odot}$/pc$^{-2}$. Kuijken and Gilmore then estimated the relative
contributions of the disk and the halo to the total integral surface density,
and concluded that the disk surface mass density in the solar neighborhood
is about 48 $\pm$ 9 M$_{\odot}$/pc$^{-2}$. The surface density
 of the identified disk matter is about 48 $\pm$ 8 M$_{\odot}$/pc$^{-2}$
which led Kuijken \& Gilmore (1991) to the conclusion that there remains no evidence
for any significant unindentified mass in the Galactic disk.
This conclusion of Kuijken \& Gilmore was disputed by Bahcall et al. (1992).
The latter authors used the method suggested in Bahcall's (1984) paper to treat non-isothermal 
stellar disk populations as a linear combination of self-consistent isothermal 
distributions. With the use of this method, Bahcall et al. (1992) found that
their sample of 125 K-giants provides $\sim$ 1.5 $\sigma$ evidence for 
dark matter in the Galactic disk.

In this paper we re-address the question of the mass density estimate
of the Galactic disk in the solar neighborhood. Determination of the integral surface density
of the disk using tracer stellar
populations involves fewer assumptions than the determination of the 
local volume density and is therefore more robust. 
In this paper we use an approach basically similar to that of Kuijken \& Gilmore (1989a).
For this investigation we choose a subsample of old red giant stars
from the Hipparcos catalog which is about 93 percent volume complete 
within $\sim \pm$ 0.4 kpc of the Sun.  Our study is not the first attempt
to use the Hipparcos data to estimate the mass density in the Galactic
disk. Pham (1997), Cr\'{e}z\'{e} et al. (1998) and Holmberg \& Flynn (2000) also used
Hipparcos data to estimate the local density of matter in
the solar neighborhood. The new element in our study is the use of a
kinematically unbiased subsample of red giants from the Hipparcos catalog
with measured radial velocities. Binney et al. (1997) have analyzed
the proper motion distribution of 1072 stars with known radial velocities
and concluded that stars with measured radial velocities form a kinematically
biased subsample. The radial velocities were discarded therefore from their
kinematical studies. 

Recently Barbier-Brossat \& Figon (2000) published a catalog of mean radial velocities for Galactic
stars which contains 20,547 new radial velocity measurements and 36,145 stellar
radial velocity measurements in total. We demonstrate that
with the catalog of Barbier-Brossat \& Figon (2000)
and the Hipparcos catalog a kinematically {\it unbiased} subsample of red giants can be extracted.
Such a subsample has measured 3-D velocities and is used in the present study for the analysis
of the kinematics of stars in the solar neighborhood.
Use of the volume-complete sample of red giants at distances extending $\sim$ 0.4 kpc
from the Sun, in combination with the kinematically unbiased subsample of these stars, allows
us to make a robust estimate of the integral surface mass density of the Milky Way disk
in the solar neighborhood. We find that the integral surface density of all gravitating
matter increases from 10 M${\odot}$/pc$^2$  within $\pm$ 50 pc  up to 42 M${\odot}$/pc$^2$ 
within $\pm$ 350 pc of the mid-plane of the disk of the Milky Way.
The last value is close to the lower limit for the surface density estimate obtained by 
Kuijken \& Gilmore (1991). An estimated volume density of the Galactic disk within
$\pm$ 50 pc is about 0.1 M${\odot}$/pc$^3$ which is close to the estimated volume density
of baryonic matter in the solar neighborhood (Holmberg \& Flynn 2000). 

\section{The Theory}

\subsection{The surface density of gravitating matter}

The total surface density of all gravitating matter can be determined from the
Poisson equation once the strength of the gravitational field ${\bf F}$ is estimated
(Binney \& Merrifield 1998):

\be
   {1 \over R} {\partial \over \partial R} \big(R F_R\big) + {\partial F_z \over \partial z} = -4\pi G \rho
   \label{eq1}
\ee

Here $R$ and $z$ are the Galactocentric cylindrical coordinates, and $\rho$ is the total volume density.   
Taking into account that the circular speed is given by the relation $v_c^2/R = - F_R$,
equation (1) can be re-written as:

\be
  \rho (R,z) = - {1 \over 4\pi G} \Big( {\partial F_z \over \partial z} 
     - {1 \over R} {\partial v_c^2 \over \partial R} \Big)
  \label{eq2}
\ee

In equilibrium, the rotation velocity of the disk does not depend on $z$ for $z \ll R$,
and the surface density of gravitating
matter within distance $z_{out}$ of the plane can be expressed as:

\be
 \Sigma_{out}(z_{out}) = 2 \int_0^{z_{out}} \rho (R,z) dz = - {F_z(R,z_{out}) \over 2\pi G}  +  
               { 2 z_{out} (B^2 -A^2) \over 2\pi G} 
 \label{eq3}
\ee

Here $A$ and $B$ are the Oort constants, for which we use the values derived by
Olling \& Dehnen (2003), $A = 15.9 \pm 1.2$ km/s/kpc and $B = -16.9 \pm 4.6$ km/s/kpc,  
$A-B = 32.8 \pm 4.5$  km/s/kpc. 
To estimate the value of $F_z(R,z_{out})$ in the first term of expression (3),
we consider the Jeans equation for the relaxed population of the 'test' stars $\rho_i$ which
is assumed to be in equilibrium with the total gravitating field $F_z$ of the disk:

\be
    {\partial (\rho_i \overline {v_z^2}) \over \partial z} + {\rho_i \overline {v_R v_z} \over R} +
    {\partial \over \partial R} \Big( \rho_i \overline {v_R v_z} \Big) =
    \rho_i F_z
    \label{eq4}
\ee
Expressing the gravitational force from Jeans Equation (4) and substituting it into
Equation (3), we get for the surface density:

$$
 \Sigma_{out}(z_{out}) = - { 1 \over  2\pi G } 
 \Big( \overline {v_z^2} {1 \over \rho_i} {\partial \rho_i \over \partial z} 
 + {\partial \overline {v_z^2} \over \partial z} \Big)\Big{\vert}_{z_{out}}
$$
\be
 - {\overline {v_R v_z} \over 2\pi G }\Big({1 \over R} + {1\over \rho_i} {\partial \rho_i \over \partial R} \Big)
   \Big{\vert}_{R_{\odot}}
-{ 1 \over 2\pi G } {\partial \overline {v_R v_z} \over \partial R} 
   \Big{\vert}_{R_{\odot}}
 + { 2 z_{out} (B^2 -A^2) \over 2\pi G}  
   \label{eq5}
\ee

As can be seen from Equation (5), the total surface density of gravitating matter in the 
solar neighborhood
can be determined by estimating four parameters for the subsample of test stars:
the vertical velocity dispersion $\overline {v_z^2}$, the vertical scale height, the cross - 
term of the velocity dispersions $\overline {v_R v_z}$, and the radial scale length of the test stars' distribution.
We show that our test sample of stars is nearly isothermal, and that a term involving
a derivative of the vertical velocity dispersion can be neglected as well as the terms
with the non-diagonal velocity dispersion components. 
We can finally write the expression for the surface density of all gravitating matter
in the disk within $\pm z_{out}$ as:

\be
 \Sigma_{out}(z_{out}) = - {\overline {v_z^2} \over  2\pi G }
 \Big( {1 \over \rho_i} {\partial \rho_i \over \partial z} \Big)\Big{\vert}_{z_{out}}
   + { 2 z_{out} (B^2 -A^2) \over 2\pi G}
   \label{eq6}
\ee

Equation (6) allows us in principle to estimate the surface density of disk gravitating matter
within $\pm z_{out}$ including the combined contribution from the thick disk and the halo.
If a sample of stars is limited in extent such that the size of the volume of study 
$z_{out}$ is small compared to the typical scale height of the thick disk/halo 
vertical density distribution, then 
$\partial (\Phi_H / \partial z) \vert_{z_{out}} \approx 0$ 
close to the plane of symmetry of the thick disk and the halo. The 
potentials of the thick disk and of the halo will then not participate in the 
vertical equilibrium of the disk. 
In such a case, Equation (6) gives an estimate of the surface density of the 
thin disk itself.

\subsection{$Z$ - distribution of the test stars}

Let's consider a simplifying situation in which all the gravitating matter in the disk 
can be assigned an average velocity dispersion $ \sigma_g $
and volume density $\rho_g(z)$.
The test sample of stars, which is assumed to be in equilibrium with the external
gravitational field of the disk, has its own
velocity dispersion  $ \sigma_i \equiv  ({\overline {v_z^2}})^{\onehalf}$
and volume density $\rho_i(z)$. The velocity dispersion of the test sample
of stars is not necessarily equal to the velocity dispersion
of the underlying gravitating matter. However by measuring the distribution
of the test stars and their velocity dispersion one can obtain information about the velocity dispersion,
and the distribution of the underlying gravitating matter in the disk.

By neglecting the  terms involving ${\overline  {v_R v_z}}$ in Equation (4),
the validity of which will be shown in our kinematical analysis below,
and by taking into account the equilibrium of the self-gravitating disk 
in the vertical direction, we can write the  equilibrium condition for a relaxed test
sample of stars:
\be
   {1 \over \rho_i} {\partial \rho_i \over \partial z} =
    {\sigma_g^2 \over \sigma_i^2}{1 \over \rho_g} {\partial \rho_g \over \partial z}
    \label{eq7}
\ee
Substituting  Spitzer's (1942) solution for the
volume density distribution in a self-gravitating slab into Equation (7), 
we can express the vertical distribution of the test stars as:
\be
  \rho_i(z) = \rho_0  \cosh^{- 2{\sigma_g^2 \over \sigma_i^2}} (z/z_0)
  \label{eq8}
\ee
where $z_0$ is the vertical scale height of the gravitating matter in the disk.
A distribution of the form given by Equation (8) was suggested by
van der Kruit (1988) for mathematical convenience. It turns out, however,
that such distributions have a simple physical meaning and
describe an equilibrium distribution of a sample of stars
in the gravitational potential of a self-gravitating
isothermal slab.

By estimating the power index $\gamma = 2\sigma_g^2/\sigma_i^2$ and the scale height $z_0$ of the spatial 
distribution sech$^{\gamma} (z/z_0)$ of the test sample of stars,
one can find in principle the average velocity dispersion and the scale height
of the gravitating matter. We find, however, that the parameters $\gamma$ and $z_0$
are highly correlated with correlation coefficient about 0.99, rendering 
it difficult to make a reliable estimate of the
velocity dispersion and the scale height of gravitating matter in the disk based on our
data samples.  

It is clear from Equation (8) that the vertical distribution of 
a test stellar population has a different functional form depending
on the velocity dispersion of the sample. The stellar populations that have velocity
dispersions higher than the average velocity dispersion of gravitating matter in the disk
will have a near-exponential distribution for $z > z_0$, while populations with velocity
dispersions close to the average velocity dispersion of gravitating matter in the disk
can be better represented by a sech$^2$ function.     
However, the scale height $\partial log(\rho_i) / \partial z $
of the distribution of sample stars can be estimated quite robustly
which allows one to estimate the surface density of gravitating
matter in the disk. 

The surface density estimate depends as well on the radial scale length of 
the distribution
of stars in the Galactic disk. We accept the value $h_R = 3.3$ kpc for the radial scale length
of the stellar distribution found from the analysis of the 2MASS survey by
L\'{o}pez-Corredoira et al. (2002). As stressed earlier, it is the vertical
scale height of the distribution of the sample stars which is more important 
for an estimate of the surface density of
gravitating matter in the disk.

\section{The Samples}

To determine the surface density of the Milky Way disk,
one needs to measure the spatial distribution of the test stars
together with the velocity dispersion of a kinematically unbiased sample.
A test stellar population should satisfy a number of criteria (see e.g., Cr\'{e}z\'{e} et al. 1998).
It should be homogeneous, i.e. the selection criterion has to be independent
of velocity and distance for the stars matching the tracer definition. 
The stellar test population has to be dense enough and occupy a sufficiently large volume.
The test population of stars must
be old in order to have had time to settle into equilibrium with the 
Galactic potential.
And the observed characteristics of the sample must be corrected for known systematic errors.

To satisfy these criteria, we choose as our basic sample of study
the bright red giant stars from the Hipparcos catalog which are older than $\sim$ 3 Gyr
assuming that all the stars have solar metallicity.
Thus, the stars in the sample
have had enough time to settle in the gravitational potential of the Milky Way.
We designate this sample RG2.
The selection criteria we use allow us to choose essentially {\it all} old red giants
within the volume of study and hence the disk properties
can be tested reliably.  For the sake of comparison, we also select another 
sample from the Hipparcos catalog.  Specifically, we also make use 
of a sample of bright red giants with ages $\sim $ 1 - 3 Gyr (RG1 sample)
under the same assumption that all the stars in the sample have solar metallicity.
By comparing the properties of this sample with our primary RG2 sample, 
we demonstrate the
differences in their spatial distributions, which reflect their
varying stages of relaxation and isothermality.  

In the following sections we discuss the details of our selection criteria as
applied to our RG2 sample of old red giants. 
Selection of the RG1 sample was made in an analogous fashion.

\section{The Spatial Distributions of the Samples}

\subsection{The selection criteria}

The survey portion of the Hipparcos catalog was designed to be complete to a limiting visual
magnitude which would be a function of Galactic latitude, and color.
For red stars, ($B-V >$ 0.8), the completeness limit is
$V \le  7.3 + 1.1 sin|b|$ 
(ESA, 1992).  This completeness can be demonstrated by comparison to the
Tycho catalog, which is know to be complete to a fainter limit than Hipparcos.
Figure 1 illustrates this by showing the frequency of 
Hipparcos $B-V >$ 0.8 stars with visual magnitudes $V \le  7.3 + 1.1 sin|b|$
relative to the number of Tycho stars which satisfy the same criteria.
From Figure 1, the Hipparcos catalog is about 93\% complete.
Note also that the incompleteness depends only slightly
on Galactic latitude $b$ and thus will not significantly affect
the spatial distribution of our sample. 

Figure 2 shows the HR diagram for the extinction-corrected Hipparcos survey stars which
satisfy the completeness criteria for the Hipparcos catalog: $V \le  7.3 + 1.1 sin|b|$ for $B-V >$ 0.8,
and $V \le  7.9 + 1.1 sin|b|$ for $B-V <$ 0.8 (ESA, 1992). Superimposed are the Yale-Yonsei isochrones
(Yi et al. 2001) for solar metallicity stars with ages 1 Gyr, and 3 Gyr.
The median metallicity of the Milky Way thin disk is slightly below solar ($[Fe/H] \approx - 0.3$)  
with a dispersion of [Fe/H] of about 0.3 dec (e.g., Haywood MN, 2002).
A sub-solar metallicity  shifts the isochrones in Figure 2
towards blue. We make the conservative assumption
of solar metallicity for our sample stars which allows us
to choose stars which are older than $\sim$ 3 Gyr. 

We choose as our RG2 sample, those stars with $M_v < 0$ and which are to the
right of (i.e. older than) the 3-Gyr isochrone,
as indicated in Figure 2.

In order that our sample traces the density distribution
of the disk to large enough distances, we apply an
absolute magnitude cut of $M_{lim} < 0.0$. This cut-off, in combination with 
the visual magnitude limit determines the boundary of our volume of study,
which has a heliocentric radius given by the expression:

\be
  R \le 10^{0.2 (7.3 + 1.1 sin |b| - M_{lim}) + 1.0}
  \label{eq9}
\ee

The completeness limit together with the absolute magnitude cut-off allow us
to choose virtually $\it all$ the bright old red giants which are inside the volume of study in the Milky Way. 
We find 1216 stars from the Hipparcos catalog satisfying the above criteria.
Correcting for the effects of extinction, which will be discussed in the 
following section, increases this number to 1476.

Figure 3 shows the peanut-shaped volume of study and the extinction-corrected
distribution of the RG2 stars, projected onto the $y - z$ plane.
The positions of the stars are shown
in a cartesian coordinate system with $z$ pointing toward
the North Galactic Pole, and $y$ in the direction
of the Galactic rotation. As illustrated by Figure 3, the RG2 sample
effectively traces the local density distribution of old red giants 
in the disk of the Milky Way, within $\pm$ 0.4 kpc of the Sun. 

In order to properly determine the vertical scale height of the stars shown 
in Figure 3, several corrections must be applied.
First, there is a purely geometrical correction
due to the shape of the volume of study. 
We make this correction by re-scaling the number of stars in an elementary
volume between $z$ and $z + dz$ to the corresponding number of stars in 
a cylinder of radius 300 pc 
centered at the position of the Sun.
The stellar distribution also must be corrected for extinction and for the 
systematic error arising from the statistical errors in the parallax 
measurements.
We discuss these corrections in the following sections.

Similar selection criteria and corrections are used to select the RG1 sample
of red giants having ages between 1 and 3 Gyrs. 

\subsection{Extinction correction}

We apply extinction corrections using the $E(B-V)$ extinction model published 
by Chen et al. (1999).
This model is based on COBE/IRAS all sky reddening maps and uses the 'infinity'
reddening maps obtained from Schlegel et al (1998). Chen et al. (1999)
tested the all-sky reddening map of Schlegel et al. with the globular cluster database
and found that this reddening map has an accuracy of 18\% but overestimates
the absorption by a factor of 1.16. This factor has also been taken into account
in our extinction corrections.  We assume the absorption in the visual passband
is 3.2 times the $E(B-V)$ extinction value.
In this manner, we re-calculated the absolute visual magnitudes and $B-V$
colors of stars in the Hipparcos catalog, and it is these which were used
in the selection of our various samples.
Figure 4 shows the visual-absorption corrections for stars in the RG2 sample,
plotted as a function of $z$. 

It is difficult to estimate the uncertainty associated with the values of
extinction correction.  We note, however, that the net affect of applying
the extinction correction is to reduce the derived vertical scale heights of
our samples by roughly 7\%.  Presumably, the possible errors in the
extinction correction will affect our results by less than this amount.

\subsection{Distance errors}

A sample selected by a lower limit in observed parallax value suffers
from a systematic effect commonly referred to as Lutz-Kelker bias,
after Lutz and Kelker (1973).  The net effect is that the observed
parallax distribution is biased toward larger values, compared to the
true parallax distribution, due to the interaction of the random
measurement errors with the steeply sloping parallax distribution.
There is some confusion as to the exact meaning of this bias, and under
what circumstances it is to be considered, as has been recently pointed
out by Smith (2003) who notes that there is no universal Lutz-Kelker bias
of individual parallaxes. We choose to model the effect, as it relates to
our Hipparcos sample, using the procedure described below.

We are interested in the effect on the observed $z$-distribution
of our sample of stars and finding a correction for it.  
The approach we have adopted is to derive a
correction function that, when applied to the observed $z$-distribution,
will recover, approximately, the true $z$-distribution.  The correction
function and its uncertainties are calculated using a Monte-Carlo approach.
A population of synthetic stars is generated based on an assumed spatial
distribution.  An absolute magnitude is assigned to each star, randomly 
drawn from a known luminosity function.  A parallax measurement error is 
similarly assigned, from an error distribution which models that of the 
Hipparcos catalog.  From these quantities are derived the star's apparent 
magnitude, 'observed' parallax, and 'observed' absolute magnitude.
A sample is then trimmed from this population using an apparent
magnitude limit, observed absolute magnitude range and corresponding
observed distance cutoff, similar to those used to form our RG2 sample 
from the Hipparcos data.  The ratio of the sample's $z$-distribution 
based on the input 'true' parallaxes with that based on the `observed' parallaxes 
defines the correction function.

It is critical that appropriate functional forms be chosen to represent
the true spatial distribution, the true luminosity function, and the
parallax measurement errors.  For the last of these, we use a fit to the
distribution of estimated parallax errors from the Hipparcos Catalog,
for our RG2 sample.  This distribution is shown in Figure 5, along with
a Gaussian centered at $-0.85$ mas/yr, and with a dispersion of 0.15 mas/yr.
While the actual error distribution deviates from a Gaussian, the mean
and dispersion of the adopted Gaussian are consistent with those of the
true distribution and we find it is an adequate representation.

The correction is more sensitive to the exact form of the luminosity function
chosen.  
Turon Lacarrieu and Cr\'{e}z\'{e} (1977) derive the corrections in absolute
magnitude based on observed parallaxes with random errors, under the
separate assumptions of a Gaussian luminosity function and of a ''top-hat''
luminosity function.  Their results differ from those of the traditional
Lutz-Kelker (1973) formulation, by factors of two or more.  We use the
Hipparcos Catalog itself to determine the {\it observed} luminosity
function of the parent population from which our RG2 sample is drawn.
Selecting Hipparcos stars with apparent magnitudes brighter than the
$V = 8$ and redder than $B-V = 1.3$
and with observed absolute magnitude $M_v < 2$, corresponds to an
approximately volume-complete sample of red giants to a distance of about
160 pc.  The Hipparcos-observed luminosity function of these stars is
shown in Figure 6.  It is well-fit by a Gaussian of width 0.78 and
centered at $-0.22$.  We shall insist that the Monte Carlo model reproduce
an observed distribution with this same mean and dispersion.

The final component of the Monte-Carlo model which must be specified
is the spatial distribution of the stars.  Of course, this
is the very distribution for which we are seeking to determine an
unknown correction function.  Thus, as with the luminosity function,
we shall select an {\it a priori} spatial distribution that reproduces
the observed spatial distribution of the RG2 sample.
We explore two different forms for the
$z$-distribution of the RG2 sample that bracket the expected range of
underlying gravitating mass distributions; a sech$^2$ functional form,
and an exponential form.
The observed, {\it i.e.} uncorrected,
spatial distribution of the RG2 sample is reasonably well fit by a
sech$^2$ with scale height of 257 pc, and with an exponential of scale height
243 pc.  The fitting is performed over the range of $z$ we feel is most
reliable, 50 pc $< z <$ 350 pc.
As for the $x$ and $y$ spatial distributions, we assume these
to be flat, \i.e. stars are randomly distributed in $x$ and $y$.
In principle, the local density inhomogeneities can alter dynamical
surface density estimates. In the galactic disks, the spiral arms are
the density inhomogeneities which have largest scales.
Current maps predict the distance between the spiral arms near the Sun to be 2.5 kpc, 
and a distance from the Sun to the nearest Sagittarius arm of 0.9 kpc (Vallee 2002). 
To smoothen a possible influence  of the density perturbations in the spiral arms of 
the Milky Way disk one needs a volume of study with a scale of about 2.5 kpc. 
Our volume of study does not satisfy this criterion.                     
However, if the numbers given by Vallee (2002) are correct, we sample the surface density 
of the disk close to the point where the density contrast imposed by the spiral arms is 
small. 

We consider the sech$^2$ $z$-distribution case first.  Through simple trial 
and error, we find that by assuming a true sech$^2$ scale height of 290 pc,
along with a true Gaussian luminosity function with mean -0.33 and 
dispersion 0.70, we recover the Hipparcos-observed spatial distribution and luminosity
function of the RG2 sample as described above.  Specifically, 250
Monte-Carlo realizations involving on average 1400 stars each, (similar
to the RG2 sample size), yield $z$-distributions that when fit with
a sech$^2$ function exhibit an average scale height of 255 pc.  The dispersion
of the 250 scale height values is 15 pc, and this we adopt as an estimate
of the uncertainty in the scale height determination.

In the case of an exponential $z$-distribution, we find that choosing
a true scale height of 280 pc, along with an input luminosity function
centered at $-0.30$ and with Gaussian dispersion 0.70, yields on average 
the desired RG2 observed luminosity function and exponential scale height
of 243 pc.  The dispersion of the scale heights is 27 pc, for the
250 model realizations.

Knowing the input ('true') parallaxes and absolute magnitudes of the stars,
a perfect 'RG2' sample can be extracted for each realization.  This can be
compared to the 'RG2' sample extracted from the same realization but
using the 'observed' values of parallax and absolute magnitude.  For both
the true and observed samples, a generalized histogram of the 
$z$-distribution is constructed using a 20-pc wide Gaussian smoothing kernel.
The ratio of the 'true' distribution to the 'observed' one then gives
us the correction function we desire.  The mean correction functions,
over the 250 model realizations, are shown in the bottom panel of
Figure 7 for both the sech$^2$ and exponential cases.  The mean correction
function determined from the sech$^2$ model is very similar to that of
the exponential model.  We choose an average of the two and describe it
as a fifth-order polynomial over the range of $z$ values for which the
observed RG2 $z$ distribution is reliable, 50 pc $< z <$ 350 pc.  The
so-adopted correction function is shown as the bold curve in the lower panel
of Figure 7.  As one might expect, the general behavior of this function
follows that of the relative parallax error as a function of $z$ for the
actual Hipparcos RG2 sample, which is also plotted in the lower panel
of Figure 7.  In the traditional Lutz-Kelker formulation, the bias
correction depends nonlinearly on the relative parallax error.

The upper panel of Figure 7 shows the uncorrected and corrected 
$z$-distributions of the real Hipparcos RG2 sample.  Each curve is a generalized
histogram constructed using a 20-pc wide Gaussian kernel function.  The
dashed curve represents the uncorrected Hipparcos measures.  The thin
solid curve shows the $z$-distribution after correction for extinction,
as described in the previous section.  The heavy solid curve shows the resulting
$z$-distribution after correcting for both extinction and distance-error
bias, using the correction function shown in the lower panel of the Figure.
The extinction correction is the larger of the two effects, but both are
significant.  Also, both act to narrow the observed distribution.

\subsection{$z$-distributions}

As was discussed in Section 2.2,
the observed $z$-distribution and vertical velocity dispersion of a ''test
particle'' sample allows,
in principle, the determination
of the average velocity dispersion and scale height of the underlying
gravitating matter in the disk.
This can be done with a fit to the corrected observed distribution  
with the function

\be
  N (z) = N_0 sech^{\gamma}((z-z_{\odot})/z_0) 
   \label{eq10}
\ee

where the value of $\gamma = 2 \sigma_g^2/\sigma_{RG2}^2$ is the ratio of the velocity dispersions
of gravitating matter and that of the test sample,
and $z_0$ is the scale height of
the gravitating matter into which the test sample of stars has settled.
We find however that we can not use this approach to determine the parameters
of underlying gravitating matter. We noted earlier that $\gamma$ and $z_0$ are highly correlated
with correlation coefficient of 0.997 for our sample 
making it difficult to separately estimate the average velocity dispersion
and the scale height of gravitating matter in the disk.
We use instead a 'traditional' approach to model the spatial distribution 
of the RG2 sample with sech$2$, sech and exponential distributions
which are typically used to model the vertical distribution of matter
in galactic disks.
Figure 8 shows the extinction and distance-error corrected 
$z$-distribution of old red giants within $\pm$ 400 pc 
as represented by the RG2 sample (dotted line)
which has been created using a Gaussian kernel function
of 50 pc. Figure 8 shows also
the best fits to this distribution with exponential
(dashed line),  sech$^2$ and sech-distributions (thin solid lines). 
The scale height of the best fit sech$^2$ distribution is 280 pc
with an uncertainty in the scale height determination of 15 pc,
as determined in the previous section. We note that this value is
in good agreement with other recent determinations of the sech$^2$
scale height of the stellar thin disk in the solar neighborhood,
284 pc (L\'{o}pez-Corredoira et al. 2002) and 282 pc (Drimmel \& Spergel 2001).
The scale height of the RG2-sample fitted with
the sech-function gives the value 185 $\pm$ 15 pc, and the scale height
of the exponential fit is 283 $\pm$ 27 pc.

The parameter $z_{\odot}$ in Equation (10) is the displacement of the mid-plane of the
distribution of our sample of stars relative to the Sun.
We find from our fits a value of $z_{\odot} =  - 1.6$ pc with the formal error of $\pm 0.5$ pc. 
The value is small
compared to the conventionally accepted value  $z_{\odot} \approx  - 10$ pc.
If we fix the displacement of the Sun, the difference between the $sech^2$ distribution with
our best-fit displacement of $-1.6$ pc, and the fixed one becomes noticeable
when the displacement of the Sun is $ -20$ pc. We conclude therefore, that with
our sample the 'resolution' of the determination of the displacement of the Sun
is about 10 pc.
We note, however, that fixing the
value for the $z$-coordinate of the Sun $z_{\odot} = - 10$ pc, or even $z_{\odot} = - 20$ pc,
does not significantly change the derived vertical scale height of the sample distribution.

Figure 9 shows the $z$-distribution of our RG1-sample of red giants corrected in a similar way for the
extinction and for the errors in parallax measurements.
The sample does not appear to be a one-component population. We show in the next section
that the velocity dispersion of the RG1-sample varies substantially as a function of $z$,
indicating that this sample is a mixture of populations with different velocity dispersions.
We therefore discard this sample from our surface density estimates.

\section{Velocity Dispersion}

Our kinematical study is based on a subsample of Hipparcos red giant stars
which have parallaxes measured with an accuracy $ \sigma_{\pi} < 0.2 \pi$,
and radial velocities taken from the Barbier-Brossat \& Figon (2000) catalog
providing the radial velocities for 36,145 stars. We impose a limit $ \sigma_{\pi} < 0.2 \pi$
in our determination of the velocity dispersion of the test stars
to avoid large uncertainties in the velocities of individual stars.
Such an assumption can potentially affect a measurement of the velocity dispersion
of the sample. We show below that the velocity dispersion of our RG2 sample
does not depend on the $z$-cooridinate, and this assumption will
not affect our determination of the velocity dispersion of RG2-sample of stars.

Figure 10 is a comparison of the Barbier-Brossat \& Figon
absolute radial velocities with the highly accurate velocities provided
by Nidever et al (2002), (with an accuracy of 0.03 km/s), for 554 stars
in common.
Excepting a few outliers, the comparison between these measurements is 
good, and thus we are confident that the sample used in our analysis 
has both good-quality velocity and proper-motion determinations.

A kinematical study must be based on a kinematically unbiased sample of stars.
We use in our study the sample of red giants older than $\sim$ 3 Gyr, that are brighter
than $M_V$ = 2.0, and that have measured radial velocities
and well-measured parallaxes.  We designate this
sample RG2$_{RV}$ and it is selected using the same 3-Gyr isochrone curve as
before, but with $M_V < $ 2.0 absolute magnitude cut-off, and with the
constraint $\sigma_{\pi}/\pi \leq $ 0.2.  This sample consists of 1868 stars.
If the radial velocities could be measured for all the red giants,
we would obviously have a kinematically 'unbiased' sample. 
In an earlier study based on the Hipparcos catalog, Binney et al. (1997) discussed  
the possibility of using radial velocity measurements. 
They compared the distribution of proper motions for their sample of 5610 stars
 with the proper-motion distribution for 1072  stars for which the 
radial velocities where available. They found that the stars in their sample
with low proper motions are under-represented in radial velocity studies,
and the sample of Hipparcos stars with radial velocities forms a kinematically
biased sample. On this basis, the radial velocities were discarded from their studies.

We must determine if our sample of red giants with known radial velocities
is kinematically unbiased. Figure 11 shows the proper-motion distribution of the
Hipparcos old red giants (thin line) compared to the distribution after
having trimmed by parallax error
$\sigma_{\pi}$ /$\pi < $0.2 (dotted line). 
In the same Figure is shown the distribution of the Hipparcos old red giants
which are similarly trimmed by parallax error {\it and} which have 
radial velocities (thick line = RG2$_{RV}$).
The three distributions differ significantly, with the distributions trimmed
by parallax error significantly under-representing the stars with low 
proper motions. This point is illustrated in Figure 12 which shows the 
distribution of proper motions of Hipparcos stars
versus their relative parallax error $\sigma_{\pi}$ /$\pi$. 
The gray dots indicate all of the old red giants, while
the black dots indicate the parallax-error-trimmed stars 
which have measured radial velocities. 
The mean proper motion as a function of parallax error differs for the stars 
which have radial velocities (thick line) from that of the general sample of 
Hipparcos old red giants (thin line).
The relative number of stars which have measured radial velocities is higher
for the stars with larger proper motions and with smaller parallax error, i.e.,
for more nearby stars. 

Nevertheless, we contend that the RG2$_{RV}$ sample
shown in Figure 11 {\it is} kinematically unbiased. Figure 13 shows the
the same three samples' distributions in tangential velocity components, 
$ k \mu_{\alpha} / \pi$ and $ k \mu_{\delta} / \pi$ where $k=4.74$ is the
conversion factor.
The generic old red giant Hipparcos sample (dotted curve) and the sample that
has radial velocity measures (thick curve) are almost indistinguishable.
The velocity dispersions for these two samples differ less than 2\% along the
$\delta$-axis, and 4\% along the $\alpha$-axis.
Hence the sample of the Hipparcos red giants with known
radial velocities and accurately measured parallaxes is kinematically
unbiased, as judged by their physical velocities.

In order to estimate the surface density of the gravitating matter in the 
Galactic disk
we need to measure the velocity dispersion of our test sample of stars,
as well as the non-diagonal term $\overline {v_R v_z}$.
To do this, we use the RG2$_{RV}$ sample of 1868 old red giant stars
which have accurately measured parallaxes $\sigma_{\pi}$ /$\pi < $0.2
and which have measured radial velocities. As was demonstrated,
such a sample is kinematically unbiased.
Before determining the vertical velocity dispersion and 
$\overline {v_R v_z}$-term, the components of the velocities have been corrected for the rotational velocity
associated with the Galactic disk:

\be
\begin{array}{ll}
   V_l = cos(b)~[ B + A cos(2l)]~ d   \\
    V_b = -A~ d ~sin(2l)~cos(b)~sin(b)\\
    V_r = A~d~ sin(2l)~ cos^2(b)
\end{array}
 \label{eq11}
\ee

Here $l$ and $b$ are the Galactic coordinates, $A$ and $B$ are the Oort constants,
and $d$ is the distance to the star. 

The relatively robust probability-plot method (Hamaker 1978) is used to determine the 
''observed'' velocity dispersion of the RG2$_{RV}$ sample, yielding a value 
$\sigma_z =$ 16.0 km/sec.  After taking into account
the measurement errors $ \sigma^2_{int} = \sigma^2_{obs} - 1/N \sum \varepsilon^2_i $ 
(McNamara \& Sanders 1978) where $\varepsilon_i $ are the errors of measurements
for individual stars, 
the intrinsic velocity dispersion of the sample is found to be 
$\sigma_z$ = 15.8 km/sec.  However, a one-component Gaussian
distribution which has velocity dispersion 16 km/sec does not fit well
the observed distribution of the vertical velocity component.
The dotted line in Figure 14 shows the observed distribution of $z$-component
velocities, ($v_z$), for the RG2$_{RV}$ sample, compared to a Gaussian 
distribution with velocity dispersion 16 km/sec (thin solid line). 
The fit is wider over most of the range in $v_z$, which points to a possible
contamination by components with higher velocity dispersion.
We consider the value $\sigma_z$ = 15.8, therefore, as an upper estimate to the
velocity dispersion. 

A least-squares fit with a general two-component Gaussian distribution
yields a value $\sigma_z$ = 12.9 km/sec for the primary component and 
a 30\% contamination from a second component with
the velocity dispersion of 24.7 km/sec. The 30\% - contamination seems 
high relative to the 8.5 \% contamination to the old thin disk found
by Siegel et al. (2002). Yet we accept the value 12.9 km/sec
as a lower estimate to the velocity dispersion of our sample.  

The thick solid line in Figure 14 shows the best fit to the distribution of 
the RG2$_{RV}$ sample trimmed on the velocity range
$-30 < v_z < +20$ km/sec. The trimmed distribution is best fit with a 
Gaussian distribution with velocity dispersion 14.6 $\pm$ 0.3 km/sec, which after 
taking into account broadening due to the measuring errors, gives a value
for the intrinsic velocity dispersion of 14.4 km/sec.
Our estimate for the velocity dispersion of the RG2$_{RV}$ sample,
and thus the RG2 sample, is therefore 
$\sigma_z =$ 14.4 $\pm$ 0.3 km/sec.

Binaries can possibly affect an estimate of the velocity dispersion
of a sample of stars. The observations imply a binary fraction between 
0.14 and 0.5. The internal velocity dispersion for the binaries 
$\sigma_b$ is about 6 km/sec (Vogt et al. 1995; Hargreaves et al. 1996).            
The 'observed' velocity dispersion of the sample $\sigma_o$ broadened 
by the binaries is related to the velocity dispersion of the sabsample 
of binary stars $\sigma_B$ as 
$ \sigma_o^2 = (1-f)*\sigma_i^2 + f*\sigma_B^2$ where $f$ is the 
fraction of binaries, and $\sigma_B^2 = \sigma_i^2 + \sigma_b^2$ 
(Hargreaves et al. 1996).
With the extreme fraction of binaries of 0.5, the 'observed' 
velocity dispersion of 14.4 km/sec is 'inflalated' by  binaries 
from the 'intrinsic' value of 13.8 km/sec.
This will result in less than nine percent correction in our 
surface density overestimate. For a more realistic estimate of 
the effect of binaries on the velocity dispersion,
we used observational data for the radial velocity measurements 
of 40 randomly selected field K giants (Harris and McClure 1983). 
The 'effective' velocity dispersion associated with the binaries 
in this sample is about 1.7 km/sec which would result in 
about 0.1 km/sec correction of the intrinsic velocity dispersion,
of our sample, or in about 1.5 
We note that the correction can larger for a sample with a smaller velocity
dispersion.

The value of the non-diagonal dispersion term can be calculated for the RG2$_{RV}$ 
sample and is found to be negligible, 
$\overline {v_R v_z} =  5.8 \pm 9.3$ (km/sec)$^2$ which results in 0.2 $\pm$ 0.3 M$_{\odot}$/pc$^2$
correction in the surface density estimate.
On this basis, the terms involving ${\overline {v_R v_z}}$ 
have been neglected in our surface density estimates. 

Strictly speaking, any sample of stars is not a single isothermal distribution
due to a spread of ages of the stars and the age-velocity relation. This might result
in a dependence of the velocity dispersion on coordinate $z$.
We have investigated this possibility with the results shown in Table 1.
We find that the vertical velocity dispersion of our RG2-sample
does not vary, in any statistically significant manner,  with $z$.
This is actually not surprising as the velocity dispersion of disk stars with ages between 2 and 10 Gyr
tends to be independent of age (e.g., Freeman \& Bland-Hawthorn 2002). 
Contrary to RG2, the velocity dispersion of the RG1 sample varies 
significantly with $z$, which shows indeed that this sample is
a mixture of populations with different velocity dispersions, and can not be used
for the dynamical estimates of the disk surface density.   

\section{Disk Surface Density}

As was demonstrated in Section 2, the vertical velocity dispersion $\sigma_z$
and the vertical scale height of the sample stars $(1/ \rho_i)(\partial \rho_i /\partial z)$
allow us to estimate the surface density of the disk's gravitating matter.
The scale heights (i.e. logarithmic derivatives) can be determined directly
from the $z$-distributions of each test sample.
Figure 15 shows the numerically determined scale height function for 
the RG2 sample (thick solid line).
The scale height of the RG2 sample is scaled
by the square of its velocity dispersion,
 $\sigma_z^2$, and thus shows the
$z$-dependence of surface density of gravitating matter in the disk.
Also shown are the smooth curves corresponding to the sech$^{2}$,
sech, and exponential functions fit to the RG2 $z$-distribution which were shown in Figure 8.
The vertical lines are error bars in the surface density determinations
which arise from the uncertainties in the scale height determinations,
and from the uncertainty in the velocity dispersion of the RG2 sample.

The surface density of
gravitating matter in the disk as 'seen' by the RG2 sample of stars
increases with $z$ up to $|z| \sim$ 400 pc. The surface density
approximately follows the sech$^2$ or sech functional distributions, and 
is inconsistent with an exponential density profile perpendicular
to the Galactic disk.
We are now in position to use Equation (6) to determine the surface
density of gravitating matter within $\pm$ 350 pc of the disk.
Using the aforementioned values of the Oort's constants,
one gets $B^2 - A^2 = 33 \pm 8$ km$^2$/sec$^2$/kpc$^2$, and the second term 
of Equation (6) evaluates
to roughly $0.1 \pm 0.02$ M$_{\odot}$/pc$^2$ at $|z| \sim$ 50 pc, and 
$0.8 \pm 0.2 $ M$_{\odot}$/pc$^2$ at $|z| \sim$ 350 pc.
Combining these with the surface density estimate shown in Figure 15 yields values for
the surface density of the Galactic disk within $\pm$ 50 pc
of 10.5 $\pm$ 0.5 M$_{\odot}$/pc$^2$, and 42 $\pm$ 6 M$_{\odot}$/pc$^2$
within $\pm$ 350 pc. For larger $|z|$,  surface density estimates
based on the RG2 sample are unreliable.

In the following section we discuss the relation of this determination 
of the surface density to the observed volume density in the solar 
neighborhood, and to dynamical estimates of the mass volume density.

\section{Discussion}

\subsection{Volume density versus surface density}

As pointed out by Kuijken \& Gilmore (1989a), a measurement of the vertical force
of the gravitational field of the disk $F_z$ is directly related to the surface density
integrated to that height. On the other hand, the vertical force  
can be determined by measuring the vertical gradient of the 
distribution of a test sample at a corresponding height.
The surface density can thus be estimated from the high-$z$ data alone without
having to worry about the detailed shape of the distribution of the test sample of stars 
near the Galactic plane.

Measuring the disk surface density at an arbitrary height $z$ allows one in principle 
to estimate the volume density distribution in the Galactic disk.
Such an estimate is based on the derivative of the surface density of the disk considered as a function of
$z$:
\be
  \rho (z) = { d \Sigma_{out} (z) \over d z}
   \label{eq12}
\ee   
The volume density estimate involves thus the second derivative of the vertical distribution of the sample stars.
The same is true for the determination of the local volume density
based on the evaluation of the Galactic potential
from the vertical density distribution and the velocity distribution of the 
sample. This approach is based on the determination of the disk gravitational
potential  $\Phi(z)$ from the integral equation  (see e.g., Fuchs \& Wielen 1993):

\be
  \rho_i (z) = 2 \int^{\infty}_{\sqrt {2 \Phi}} 
  {f(|\omega_0|) \omega_0 d \omega_0 \over
 \sqrt{\omega_0^2 - 2 \Phi} }
  {\mbox ,}
  \label{eq13}
\ee
which can be evaluated if the vertical spatial distribution $\rho_i (z)$ and the velocity distribution
$f(\omega_0)$ of the test sample are known.  The local dynamical density can be determined then from the Poisson equation:

\be
  \rho = {1 \over 4 \pi G} \Big( {d^2 \Phi \over dz^2} \Big)
  \label{eq14}
\ee

The estimate of the local volume density involves thus a knowledge of the second
derivative of the distribution of the test stars close to the mid-plane of the disk
which makes the volume density estimates less robust compared to 
surface density measurements. This point is clearly illustrated in Figure 15.
The first derivative of the spatial distribution of the sample
is somewhat uncertain due to uncertainties in the   
extinction correction. 
An estimate of the local volume density close to the mid-plane of the disk would
therefore be less robust than the surface density determination.
We note however that the local volume density estimated from our
surface density measurements is close to recent volume density measurements
obtained by other groups (see, e.g. Holmberg \& Flynn 2000).


Nevertheless, our results allow us to make a rough estimate of the volume density of gravitating
matter in the Galactic disk. We find that the surface density of gravitating matter in the disk
is about 10.5 $\pm$ 0.5 M$_{\odot}$ / pc$^2$ within $\pm$ 50 pc.
This gives
a value for the volume density of gravitating matter of about $\sim$ 0.105 $\pm$ 0.005 M$_{\odot}$/pc$^3$ 
under the conservative assumption that the gravitating matter is distributed homogeneously. 
This value should be compared to the estimated volume density of visible disk matter
0.095 M$_{\odot}$/pc$^3$ (Holmberg \& Flynn 2000). Our dynamical estimate of the volume 
density of the Galactic disk is thus well comparable with the identified matter in the solar neighborhood,
and, at the 1 - $\sigma$ level, is 5 -- 20 percent larger than the volume density of identified matter.
If, however, the volume density is distributed non-homogeneously within $\pm$ 50 pc,
this would lead to a larger discrepancy between the observed, and the dynamical volume density
estimate close to the mid-plane of the disk. We concur, however, with the conclusion
of Holmberg \& Flynn (2000) that there is no compelling evidence for a significant
amount of dark matter in the disk.

\subsection{The thickness of the gas layer}

Olling (1995) and Olling \& Merrifield (2001) discussed the constraints on the local surface density
based on a theoretical estimate of the thickness of a gas layer settled into
equilibrium within the gravitational field of the Galactic disk.
Their estimate gives a thickness of about 500 pc for the HI layer settled in the solar neighborhood within
a self-gravitating disk, assuming that the potential is dominated by an isothermal stellar disk.
This value is larger than the observed thickness of the
atomic hydrogen layer in the solar neighborhood of 410 $\pm$ 30 pc. Olling \& Merrifield (2001)
conclude therefore that a significant contribution to the total gas pressure from cosmic rays
and magnetic fields can be ruled out.

Using the same assumption that the potential is dominated by a self-gravitating  disk with sech$^2$
scale height of 280 pc, and a total surface density of 48 M$_{\odot}$/pc$^2$ we estimate
the thickness of the molecular hydrogen in the solar neighborhood.  
Assuming the velocity dispersion of molecular hydrogen to be 7 km/sec (Stark \& Brand 1989, Binney \& Merrifield 1998),
we estimate with the help of Equation (12) by Olling \& Merrifield (2001) that the  
FWHM of the $z$-distibution of the molecular hydrogen is about 300 pc.
This is in stark contradiction to the observed thickness of the molecular
hydrogen layer in the solar neighborhood, which is about 140 pc (Binney \& Merrifield 1998). 

The discrepancy between the observed thickness of the molecular hydrogen and estimates
based on the assumption that the gravity is dominated by the stellar component,
points at the importance of the self-gravity of the gas. 
Narayan \& Jog (2002) recently discussed the vertical scale heights of 
molecular and atomic hydrogen in the disk of the Milky Way taking into account 
gravitational coupling between the gas components
and the stellar disk. They found that the common gravity
of a stellar disk,
of the atomic hydrogen, and of the molecular hydrogen can explain the 
observed scale height
of the molecular hydrogen in the disk of the Milky Way.
We note, however, that Narayan \& Jog (2002) overpredict by about 25 \% the thickness of the molecular
hydrogen in the solar neighborhood (see their Figure 1b). The discrepancy can be larger taking into account that 
Narayan \& Jog (2002) adopted a somewhat low velocity dispersion for the
molecular hydrogen of 5 km/sec. 
The distribution of the molecular hydrogen, thus, points to a 
non-homogeneous distribution of gravitating matter within $\pm$ 70 pc of the
mid-plane of the Galactic disk, and to a larger value of the volume density of gravitating
matter in its mid-plane compared to the volume density of the observed matter.

\subsection{Exponential versus sech$^2$ distribution}
 
As we have demonstrated, the vertical distribution of an isothermal sample of test stars
is determined to first approximation by the ratio of the velocity dispersion of the sample to
the average velocity dispersion of the gravitating matter in the disk. 
If the sample has a velocity dispersion larger than that of the gravitating
matter in the disk, the distribution of the sample will be closer to exponential. 
A sample which has a vertical velocity dispersion close to that of the
underlying gravitating matter, will be distributed
according to a sech$^2$ law.
There is observational evidence for an exponential distribution
of stars perpendicular to their disk in external galaxies.
Vertical light distributions in edge-on spiral galaxies (de Grijs \& van der Kruit 1996) 
 show that
the best fitting models are either exponential or simple sech($z$) distributions.
One explanation for an exponential distribution of a star sample is that the scale height of 
the distribution of the sample stars is large compared to the scale height of the gravitating
matter, i.e. the test sample is kinematically hotter than the gravitating matter.
A near-exponential distribution of a sample can also be a result of a mixture of groups of
stars with different ages and hence with different scale heights.
We think, that this is the case with
our RG1 sample which possibly contains a considerable number of young kinematically unrelaxed
stellar populations mixed with the subgroups of older stars. A determination of age of
the stellar samples used in dynamical studies is therefore critical.

In the case of the older RG2 sample, the vertical distribution  
is better fit by sech$^2$ or sech functions rather than
an exponential distribution. This can be seen in Figure 8, and especially in Figure 15.
The spatial distribution of the RG2 sample precludes,
thus, a distribution of gravitating matter in the disk with scale height
much smaller than 280 $\pm$ 15 pc.

\subsection{Local stability criterion}

With our estimate for the value of the surface density of gravitating matter in the solar neighborhood,
we can estimate the local gravitational stability of the Galactic disk. The local stability of a one-component
collisionless disk is governed by Toomre's (1964) stability criterion $Q > 1$, where $Q$ is given by the expression:

\be
    Q = {\sigma_R \kappa \over 3.36 G \Sigma } 
  \label{eq15}
\ee

Here $\sigma_R$ is the average radial velocity dispersion (i.e. along a disk radius) 
of gravitating matter in the disk,
$\kappa$ is the epicyclic frequency, $G$ is the gravitational constant, and $\Sigma$ is the
total surface density of the Galactic disk. 

An estimate of the local stability parameter is
prone to a number of uncertainties associated with the uncertainty in knowledge
of the local surface density, the 'effective' radial velocity dispersion
in the disk, and uncertainty in determination of the epicyclic frequency $\kappa$.
Based on the Oort constants taken from Olling \& Dehnen (2003), 
the value for the epicyclic frequency $\kappa$ is about 47 km/sec/kpc. 
The value for the radial velocity dispersion is, however, much more uncertain.  
The Galactic disk is multicomponent,
and the components have different values of velocity dispersion. Stars, for instance
have radial velocity dispersions of about 20 - 35 km/sec. To estimate the local stability
of the gravitating disk from Toomre's criterion, one needs to know the effective radial velocity dispersion
of the gravitating disk.

One way to estimate the local stability of the Galactic disk would be to separate it into a number of isothermal components
and to use a criterion for instability suitable for a multi-component disk (Rafikov 2001).
Another possibility is to estimate an effective radial velocity dispersion of its gravitating matter.
Rafikov (2001) derived a stability criterion for a disk consisting of a number of components with
different radial velocity dispersions.
A ten percent admixture of a 'cold' component which has a radial velocity dispersion of 10 km/sec
leads to an 'effective' radial velocity dispersion of 22 km/sec for a collisionless disk
with radial velocity dispersion 30 km/sec. We can assume therefore that an average radial
velocity dispersion of the disk in the solar neighborhood is about 20 km/sec.
With an estimated surface density of all gravitating matter in the solar neighborhood of
42 $\pm$ 6 M$_{\odot}$/pc${^2}$,
and with the adopted value for the effective radial velocity dispersion 
of 20 km/sec, the $Q$-parameter in the solar
neighborhood is about 1.5 $\pm$ 0.2.  Such a value for the $Q$-parameter
places the Galactic disk in the solar neighborhood well above the marginal stability.
However accepting a more 'conventional' value for the epicyclic frequency of 37 km/sec/kpc,
we get a value for the local stability parameter of 1.18 $\pm$ 0.16.  
These estimates indicate that the Galactic disk is locally stable, which does not preclude
the Milky Way disk being globally unstable with a corotation radius located inside the solar circle.

\section{Conclusions}

We have used Hipparcos parallaxes and proper-motion measurements to form a kinematically
unbiased sample of red giant stars that have measured radial velocities taken from the catalog of
Barbier-Brossat \& Figon (2000). The absolute magnitude cut-off (M$_V < 0$) and Yale-Yonsei solar metallicity
isochrones allow us to form a
sample of 1476 red giants older than $\sim$ 3 Gyr, assuming their metallicity to be solar, 
which is about 93 \% complete in our volume of study.
Using the density profile for these red giants and a determination of the velocity dispersion
of a similar sample, we re-determine the surface density of the Galactic disk within $ \pm$ 0.4 kpc of the Sun.

A determination of the surface density of the Galactic disk requires measurement of the first
derivative of the spatial distribution of the test stars above the plane in combination
with the measurement of the velocity dispersion of the sample. An estimate of the
dynamical volume density near the Sun requires measurement of the second derivative
of the distribution of the test stars near the Galactic plane, where the extinction correction
uncertainties are largest. We concur therefore with the conclusion of Kuijken \& Gilmore (1989a)
that the estimates of the surface density of the Galactic disk are more robust compared to the
dynamical volume density estimates.

An estimate of the first derivative of the distribution of red giants 
together with our estimate of the intrinsic velocity dispersion of the sample of 14.4 km/sec
yields a surface density of gravitating matter in the Galactic disk that 
varies from 10.5 $\pm$ 0.5 M$_{\odot}$ / pc$^2$ within $\pm$ 50 pc to  42 $\pm$ 6 M$_{\odot}$ / pc$^2$
within $\pm$ 350 pc.

The distribution of a sample of stars in equilibrium with the gravitational
potential of a galactic disk is described by the function
sech$^{ 2(\sigma_g^2 /\sigma_i^2)} (z/z_0)$. By estimating the power index of the distribution
of the test stars and its scale height one can estimate an average velocity dispersion
and scale height of the underlying gravitating matter in the disk.
However, the scale height and the power index of the test sample distribution are highly
correlated with correlation factor of about 0.99 for our sample, 
which does not allow us to make a reliable estimate of these parameters.
The surface density estimate is however quite robust within $\pm$ 50 to $\pm$ 350 pc.

An estimate of the volume density of gravitating matter gives, at the 1-$\sigma$ level, a value
0.1 -- 0.11  M$_{\odot}$/pc$^3$ under the conservative assumption that the gravitating matter is distributed homogeneously. 
This is  5 -- 20 percent larger than the volume density of identified matter in the solar neighborhood 0.095 M$_{\odot}$/pc$^3$.
The discrepancy might be larger if the volume density of gravitating matter is distributed
non-homogeneously close to the mid-plane of the Galactic disk.
A small thickness of the molecular hydrogen layer near the Sun confirms indeed
that this might be the case. We conclude, however, that our data do not provide 
evidence for a large amount of unindentified matter in the solar neighborhood.  

Acknowledgments

We thank the referee, C. Flynn, for his thoughtful questions and comments which 
helped to improve the paper.
We wish to thank members of the Yale Department of Astronomy
for their helpful comments on an earlier version of this study.  In particular,
Professors Pierre Demarque, Robert Zinn, Richard Larson, and Sabatino Sofia
provided valuable suggestions, including the use
of isochrones to better delineate the red giant samples.
We thank also Steve Shore for  reading the manuscript and providing many
valuable comments.

This work was supported in part by grant No. AST 0098548 from the National Science Foundation.

\clearpage

\figcaption[f1.eps]{Distribution of Hipparcos red stars with
$B-V >$ 0.8 and with visual magnitudes $V \le  7.3 + 1.1 sin|b|$, compared
to that of Tycho stars satisfying the same criteria.
The Hipparcos catalog is about 93 \% complete
compared to the Tycho catalog which is effectively complete at these
magnitudes.
 \label{fig1}
}

\figcaption[f2.eps]{The HR diagram constructed from the Hipparcos catalog for the survey
stars satisfying the apparent-magnitude completeness criteria. Dotted lines show the superimposed
Yale-Yonsei isochrones for solar metallicity stars with ages 1 Gyr and 3 Gyr.
Regions
RG1 and RG2 select the old bright red giants with ages between 1 Gyr and 3 Gyr (RG1)
and older than 3 Gyr. Colors and absolute magnitudes of
stars have been corrected for extinction.  
 \label{fig2}
}

\figcaption[f3.eps]{ The projection of the RG2 sample of all Galactic old bright red giants onto the
$z - y$ plane, and a cross-section of the peanut-shaped volume of study. 
The $z$-axis points towards
the North Galactic Pole, and $y$-axis in the direction of the Galactic rotation.
The sample is deep enough to study the local density distribution in the Milky Way disk
within $\pm$ 0.4 kpc.
 \label{fig3}
}

\figcaption[f4.eps]{Calculated visual absorption corrections for each of the 
bright red giant stars in our RG2 sample
as a function of distance from the plane, $z$.
 \label{fig4}
}

\figcaption[f5.eps]{The distribution of the parallax errors of the red giants
from the Hipparcos Catalog (solid line) normalized to unity. Dashed line shows a Gaussian
distribution centered at 0.85 mas/yr with a dispersion of 0.15 mas/yr
which was used in our simulations.
\label{fig5}
}

\figcaption[f6.eps]{The Hipparcos-observed luminosity function of red giant stars
with $M_V < 2$ (solid line). The distribution is well fitted by a Gaussian
of width 0.78 centered at $-0.22$ mag. The dashed line shows the 'observed' luminosity
function reproduced from 250 Monte-Carlo simulations. The reproduced distribution
has the same mean and dispersion as the Hipparcos-observed luminosity function.  
\label{fig6}
}

\figcaption[f7.eps]{The upper panel shows the $z$-distribution of red giant stars older than $\sim$ 3 Gyr 
in the 'real' Hipparcos catalog 
(dashed line) compared to the extinction-corrected distribution of 'real' Hipparcos red giants 
(thin solid line), and that distribution adjusted for individual parallax measurement 
errors (heavy solid line). 
The lower panel shows the parallax error correction factor calculated for the exponential,
and sech$^2$ - distributions of stars together with the
relative parallax error $\sigma_{\pi}$ / $\pi$ of stars in the sample as a function of $z$.
  \label{fig7}
}

\figcaption[f8.eps]{ The extinction and distance error corrected $z$-distribution of old red
giants represented by the RG2 sample (dotted line). Also shown are the best fits to this
distribution with exponential (dashed line), sech$^2$ and sech-distributions (thin solid lines).
The spatial distribution functions are created with the use of a 50-pc Gaussian kernel function.
  \label{fig8}
}

\figcaption[f9.eps]{The vertical distribution of the RG1 sample of stars which have ages between $\sim$ 
1 and 3 Gyr (thin solid line) as
compared to the distribution of old red giants, the RG2 sample,  (thick solid line).
  \label{fig9}
}

\figcaption[f10.eps]{Distribution of differences of radial velocity measurements for 889 stars 
obtained by Nidever et al. (2002) and Barbier-Brossat \& Figon (2000).
  \label{fig10}
}

\figcaption[f11.eps]{Proper-motion distribution of all Hipparcos
red giants which are brighter than $M_v = 2.0$ and older than $\sim$ 3 Gyr (thin solid line),
 as well as after trimming by the parallax error $\sigma_{\pi}$ /$\pi < $0.2 (dashed line). 
The thick solid line shows the proper-motion distribution of the Hipparcos red giants trimmed
by parallax error and which have measured radial velocities.
  \label{fig11}
}

\figcaption[f12.eps]{Proper-motion distribution of all Hipparcos red giants brighter than $M_v = $ 2.0 (gray dots)
as a function of the relative parallax error $\sigma_{\pi}$ /$\pi$. Heavy dots show the distribution of the
Hipparcos old red giants which have measured radial velocities. The thin and the thick solid curves
show the average value of the proper motion in the samples as a function of the
parallax error for all Hipparcos old bright red giants, and for those which have measured radial
velocities, respectively.
  \label{fig12}
}

\figcaption[f13.eps]{Tangential velocity distributions, $\mu_{\alpha} k/\pi$, $\mu_{\delta} k/\pi$,
 of all Hipparcos red giants (thin solid line) and
of the Hipparcos old red giants trimmed by the 
parallax error $\sigma_{\pi}$ /$\pi < $0.2 (dashed line),
compared to the proper-motion distributions of old red giants which also have measured radial velocities
and similarly trimmed by the parallax error $\sigma_{\pi}$ /$\pi < $0.2
(thick solid line). Here $k=4.74$ is the conversion factor converting parallaxes (in mas) and
proper motions (in mas/yr) into velocities in km/s.
All trimmed proper motion distributions are similar to the unbiased distribution 
of all old red giants and are thus deemed kinematically unbiased.
  \label{fig13}
}

\figcaption[f14.eps]{Distribution of the vertical velocity of the old red giants (dotted line) 
created with a 2-km/sec Gaussian kernel function. In gray is shown the histogram
of the velocity distribution of the RG2$_{RV}$  sample plotted with a 2 km/sec binning.
These are
compared to Gaussian distributions with intrinsic velocity dispersion 16.0 km/sec (thin line)
and 14.4 km/sec (thick line) for the sample
trimmed at the velocity range $-30 < v_z < +20$ km/sec. The Gaussian with velocity dispersion 16 km/sec 
is wider than the observed distribution, indicating a possible
contamination of the distribution from components with higher velocity dispersion.
  \label{fig14}
}

\figcaption[f15.eps]{The scale height, $1/\rho_i (\partial \rho_i / \partial z)$, as a function of $z$
for the vertical distribution of the old red giant sample RG2 (thick curve).
The smooth, narrow-lined curves represent the
fits to the RG2 sample illustrated in Figure 8.
The scale heights are scaled by the square of the velocity dispersion of the sample,
and thus give a measure of the surface density of gravitating matter in the disk.
The vertical lines represent the error bars in the surface density determinations
which arise from the uncertainties in the scale-height determinations,
and from the uncertainty in the velocity dispersion of the RG2 sample.
A Gaussian kernel function of width 50 pc was applied to the observed spatial distribution, prior to
being numerically differentiated.
}
\clearpage

\begin{table}
\begin{center}
\caption{\label{tbl-1}}
\begin{tabular}{rrrrr}
\tableline\tableline
Sample   &$\mid$z$\mid$    &No.       & $\sigma_{W_{rms}}$  & $\sigma_{W_{prob}}$ \\
         &(pc)             &(stars)   & (km/s)              & (km/s)\\
\tableline\tableline
RG1  & 0 - 50    & 565 & 16.1 & 12.9  \\
     & 50 - 100  & 435 & 18.9 & 15.7   \\
     & 100 - 150 & 201 & 19.3 & 15.8   \\
     & $>$150  & 106 & 20.1 & 18.6   \\
\tableline
RG2  & 0 - 50    & 868 & 19.0 & 15.8  \\
     & 50 - 100  & 586 & 18.3 & 15.9   \\
     & 100 - 150 & 288 & 18.7 & 16.4   \\
     & $>$150   & 126 & 19.2 & 17.0   \\
\tableline
\end{tabular}
\end{center}
\end{table}

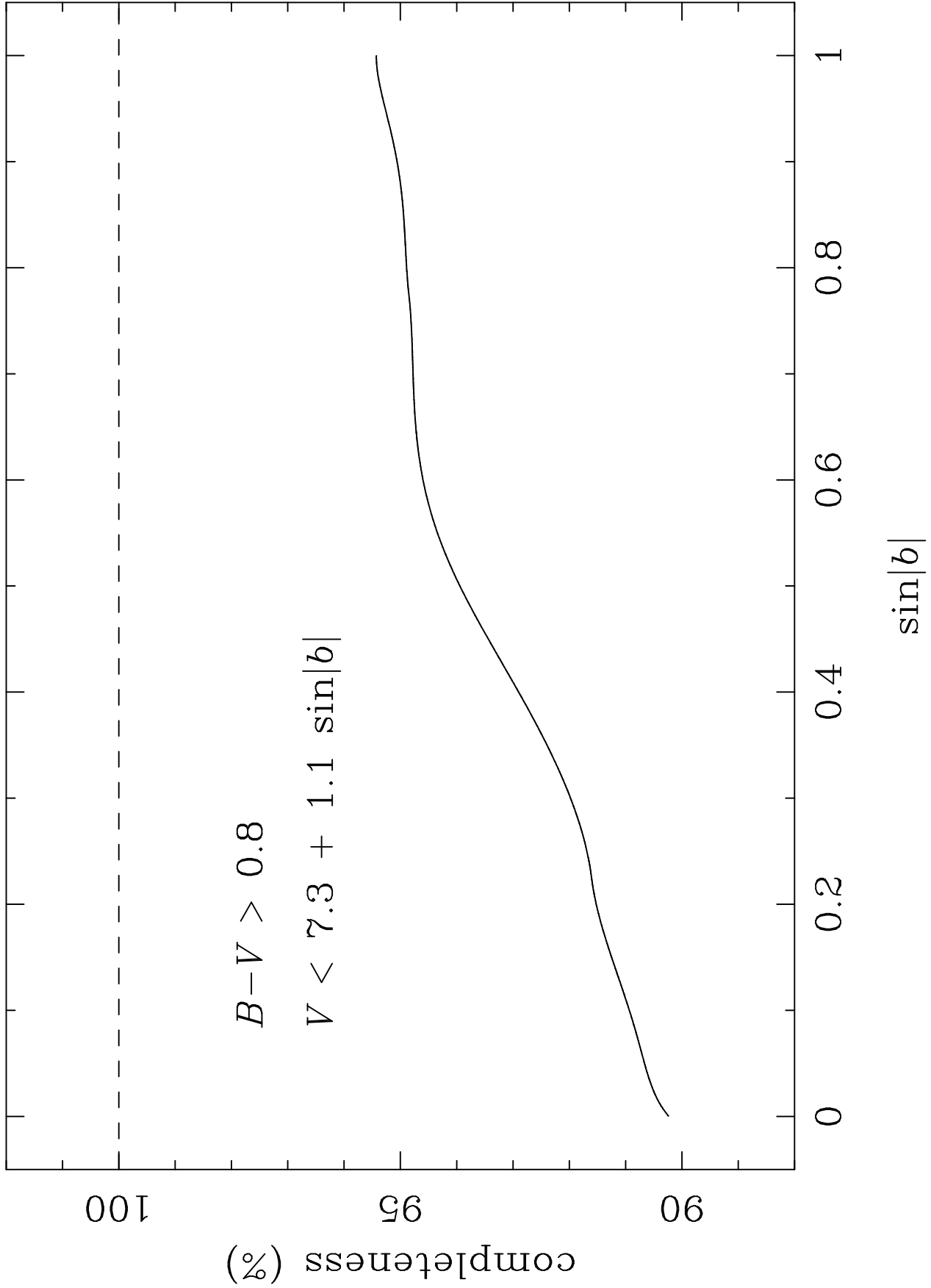
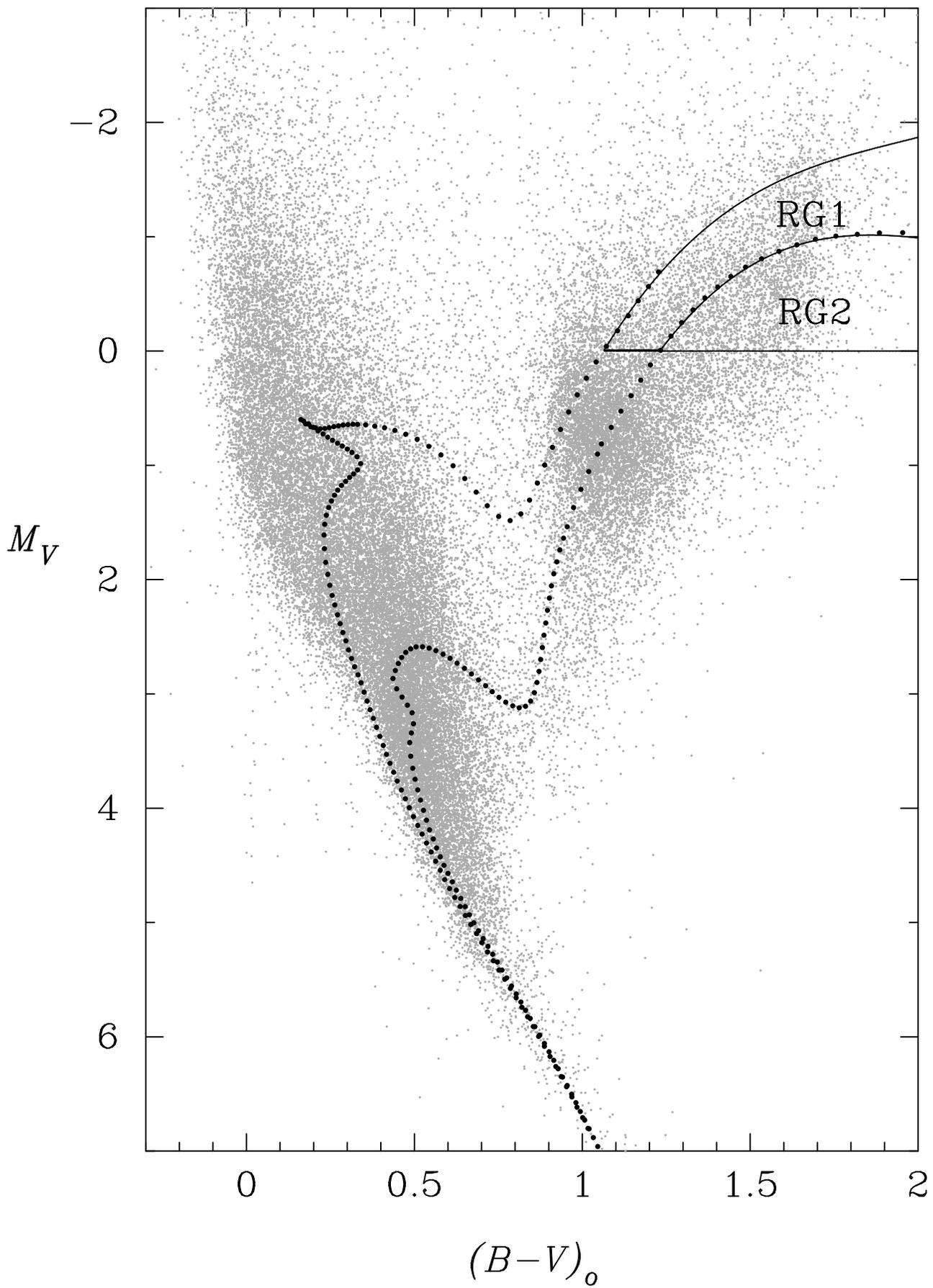
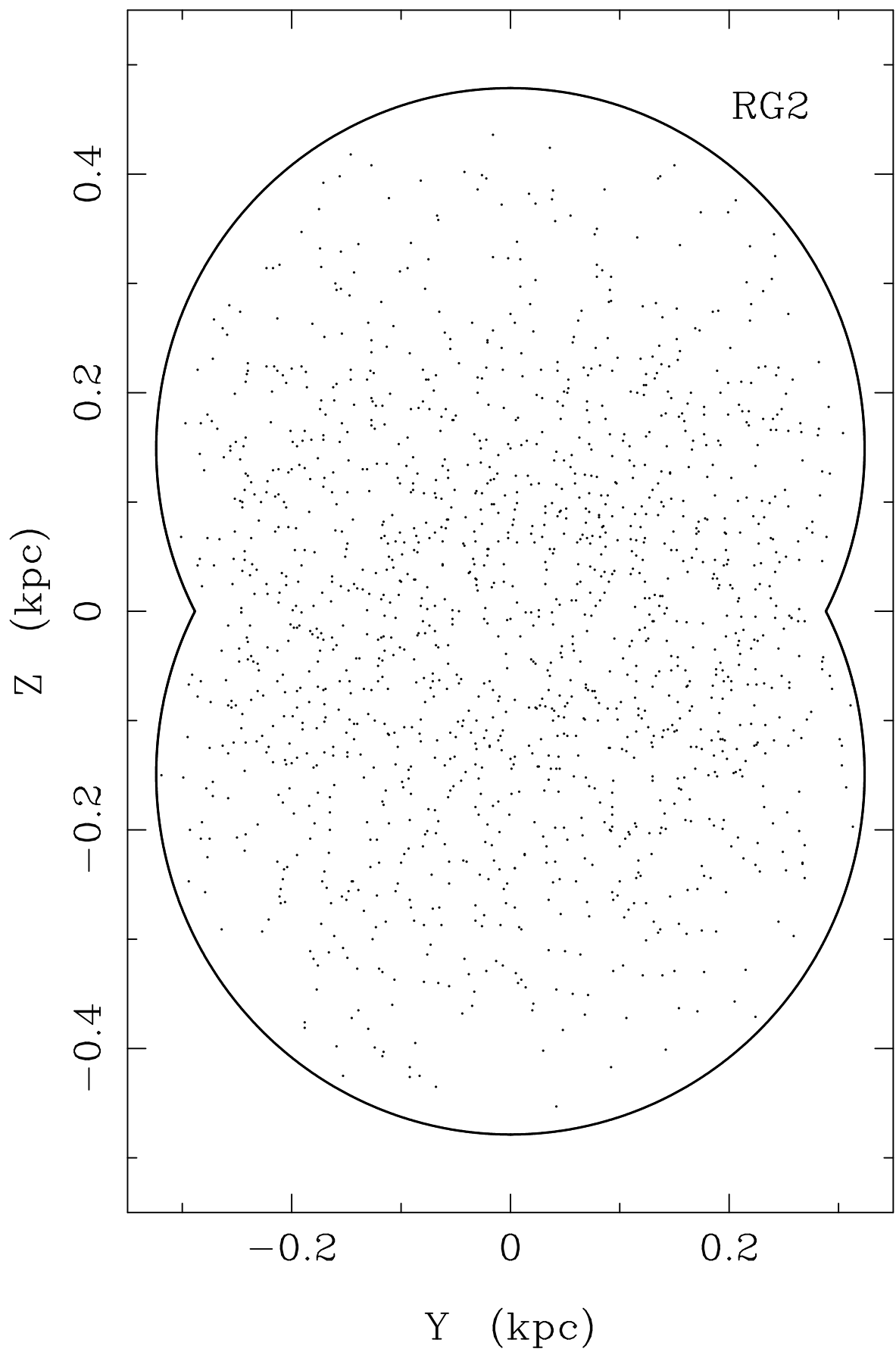
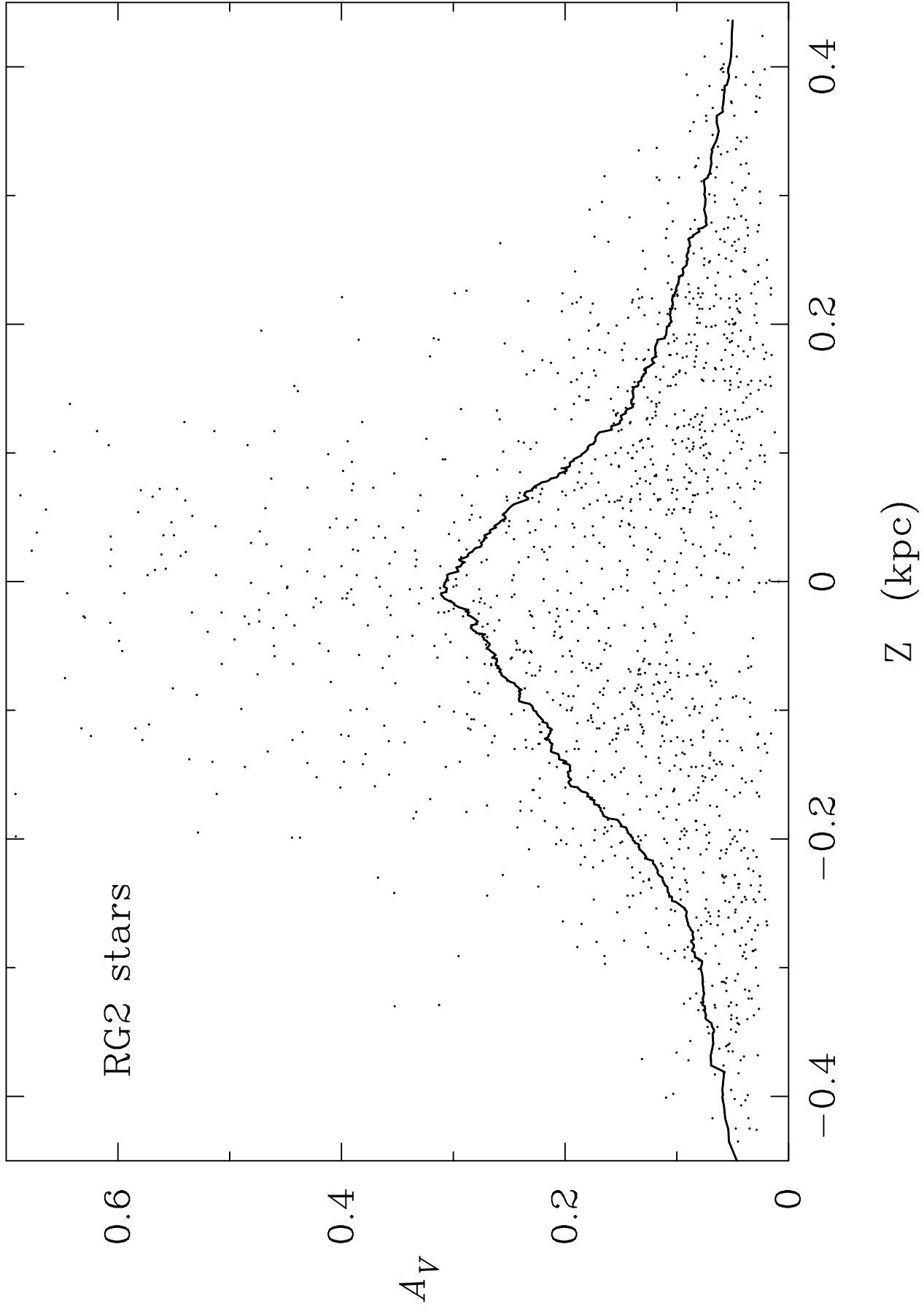
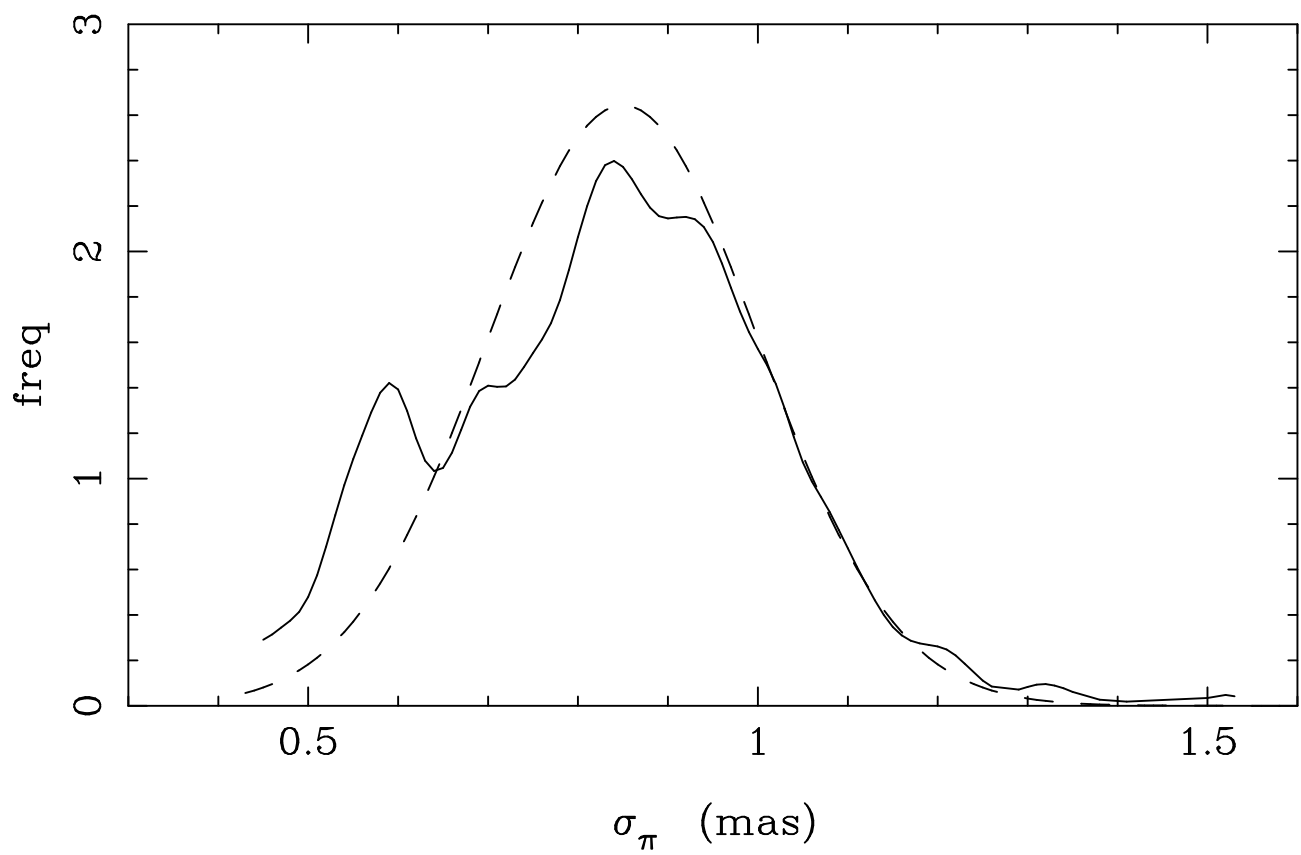
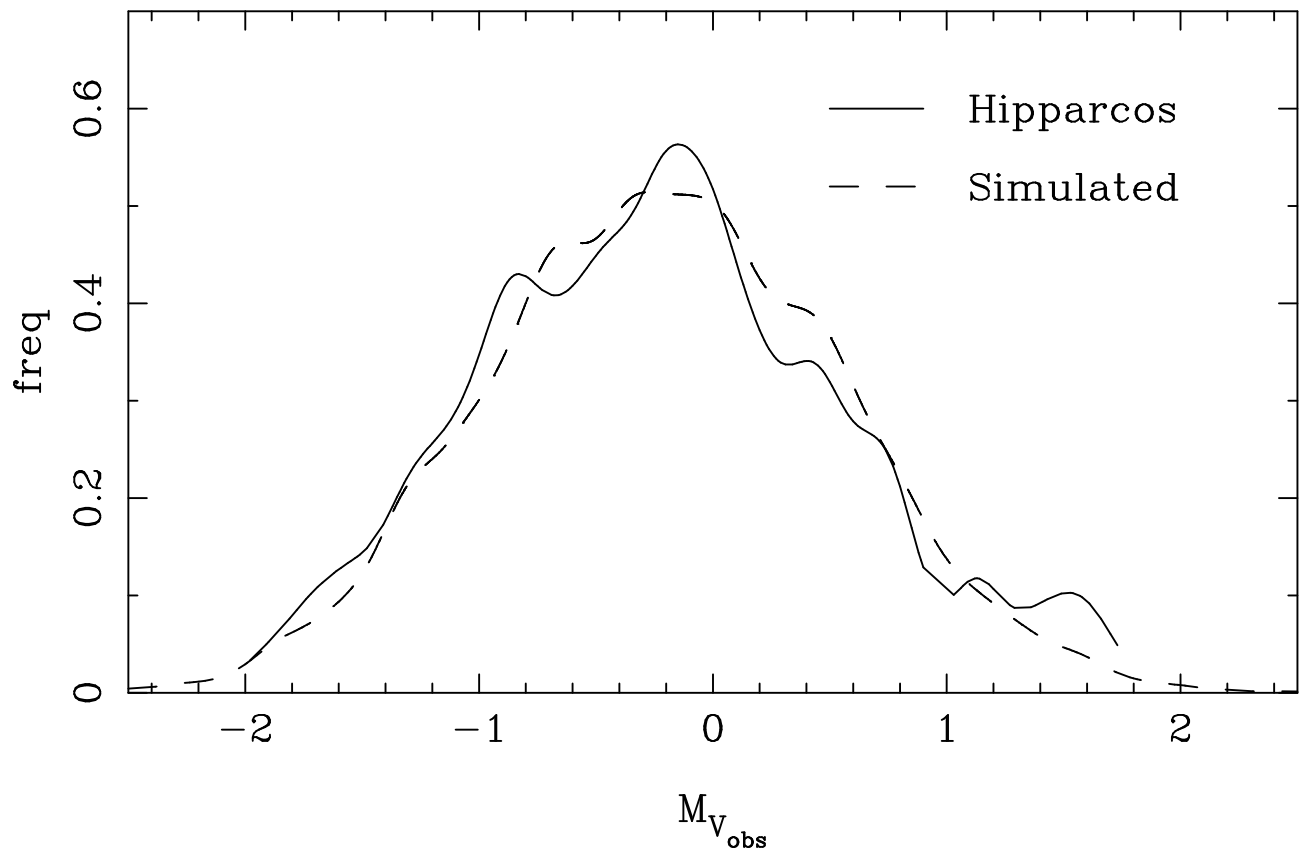
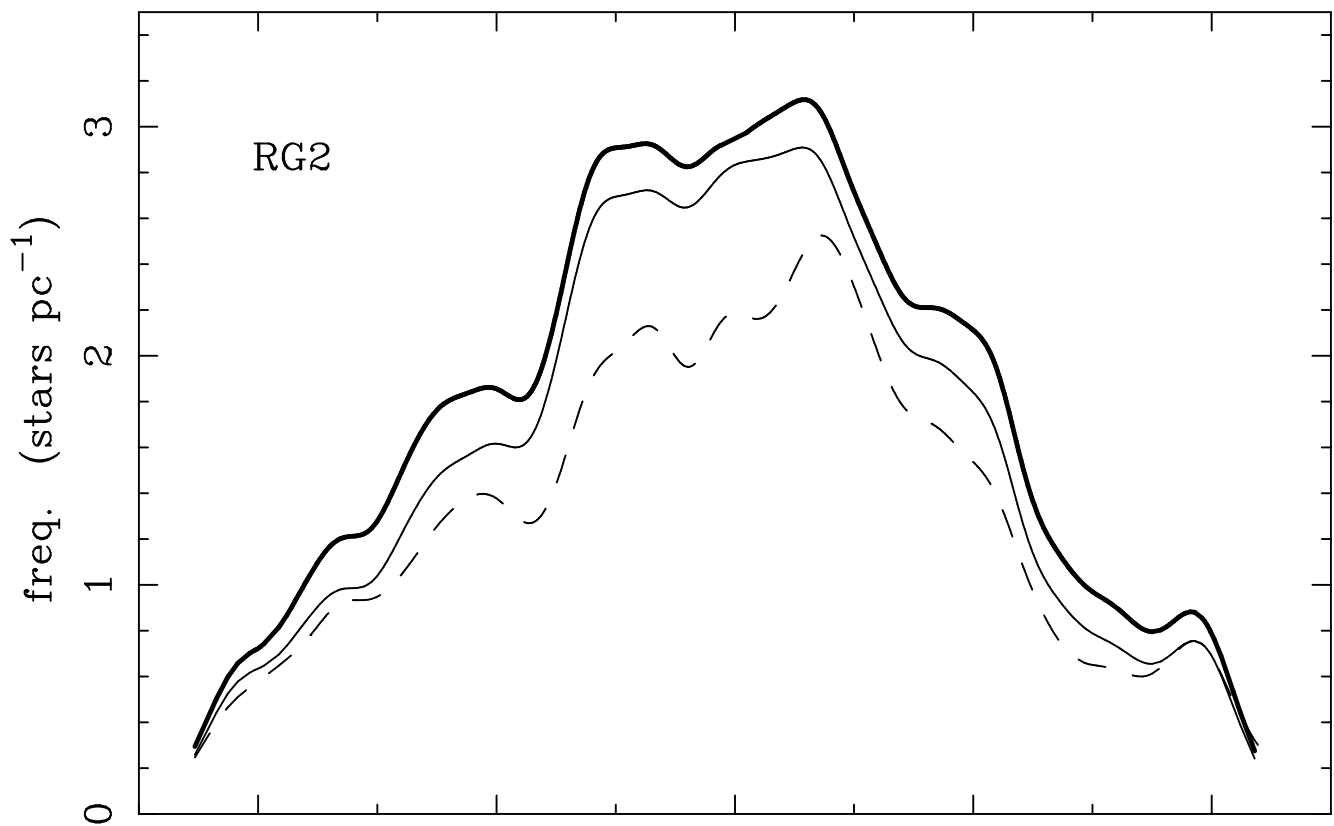
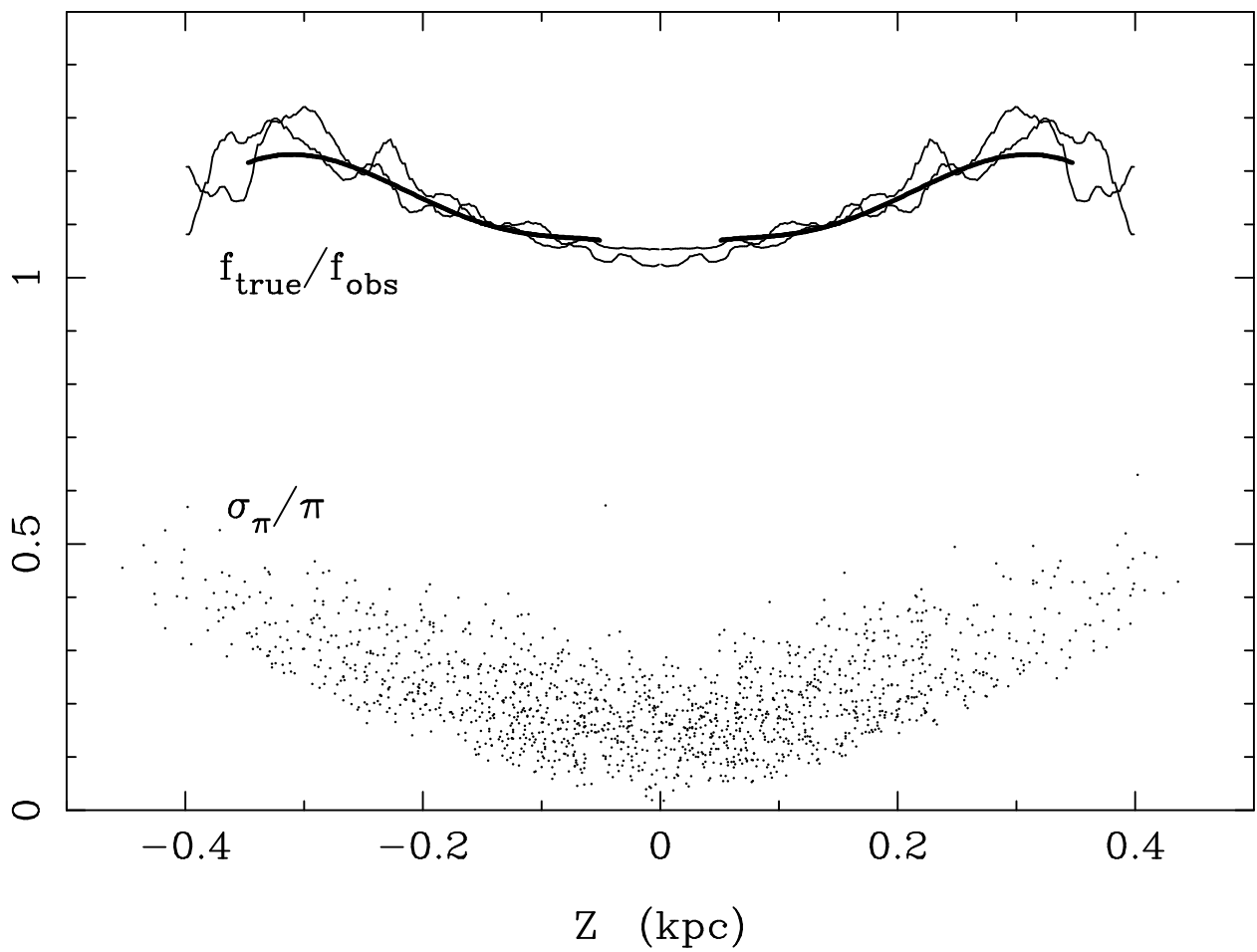
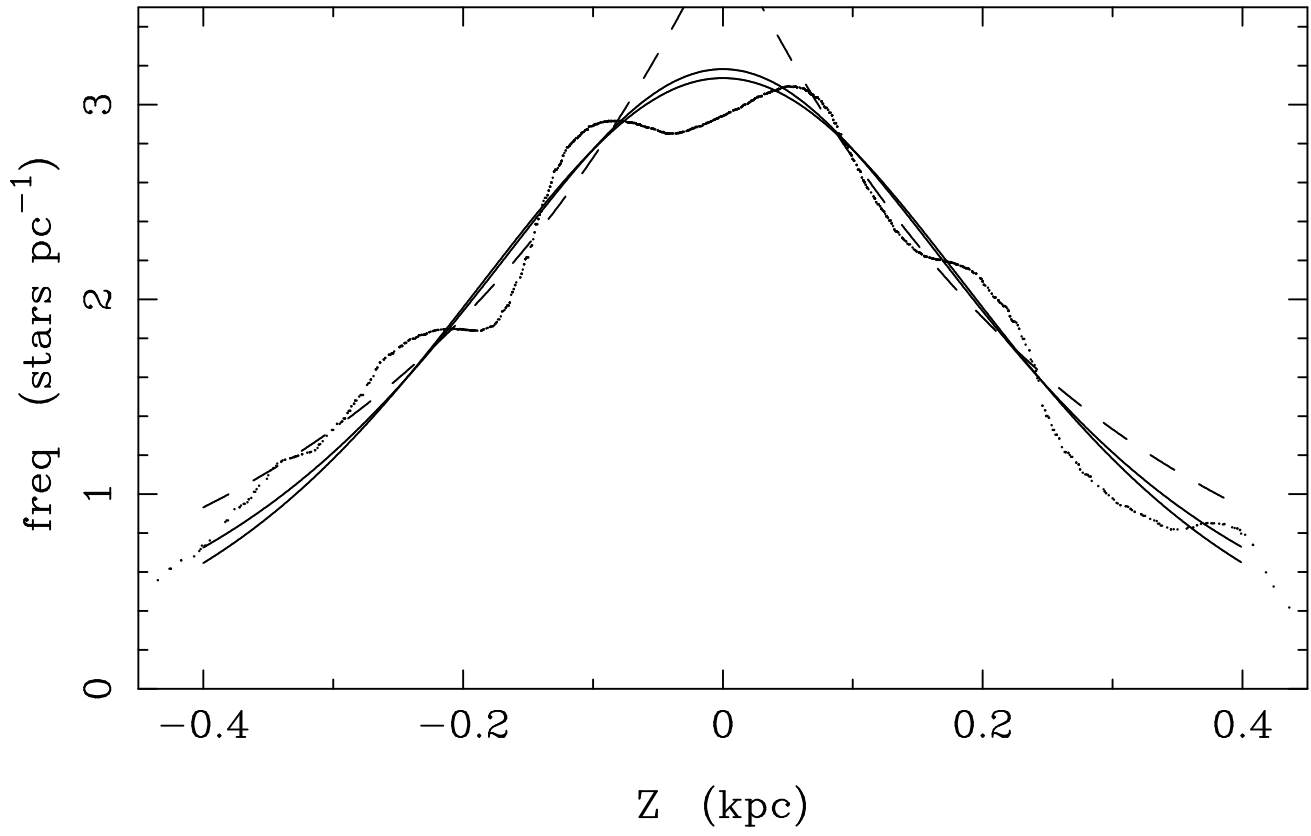
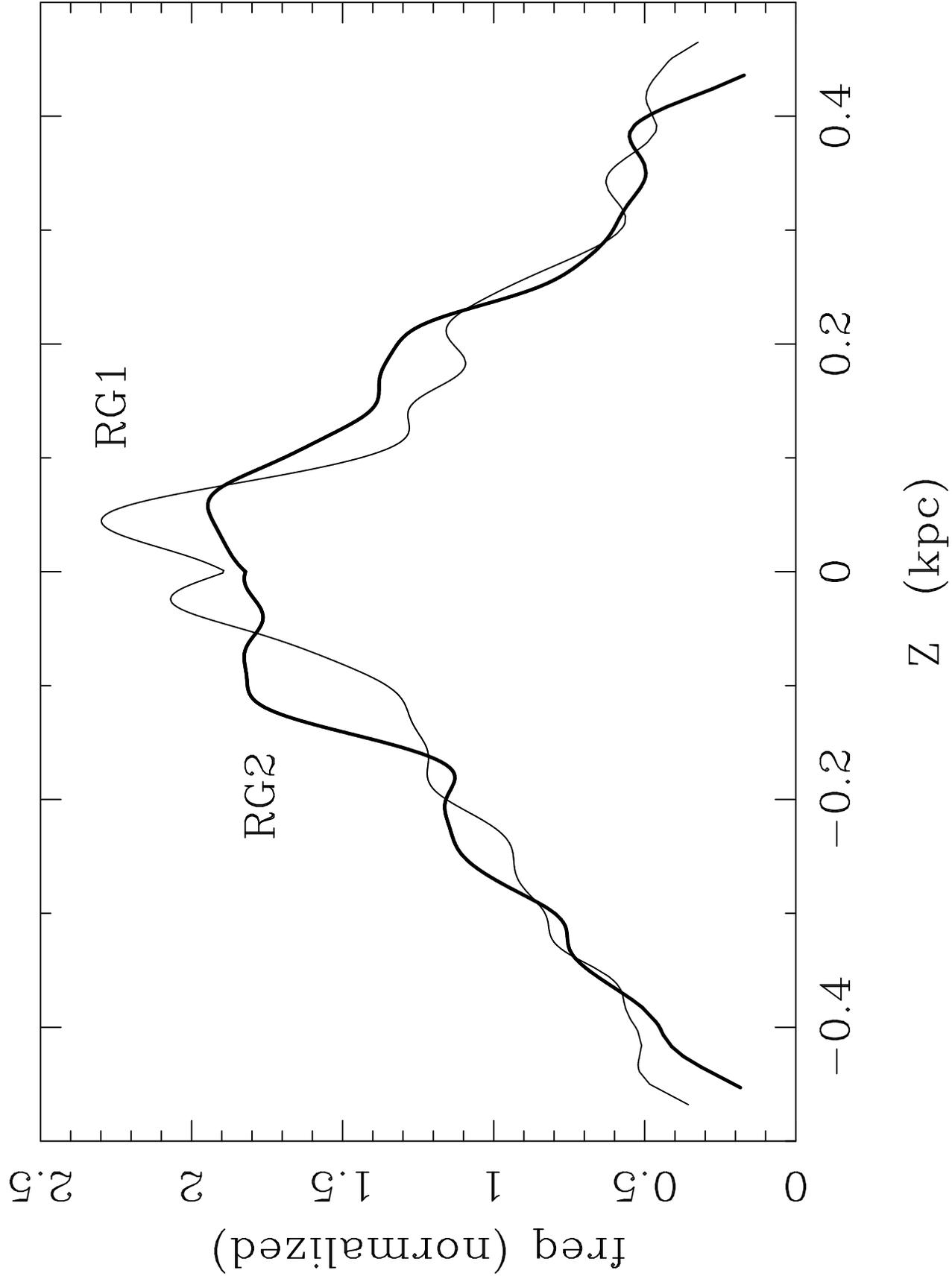
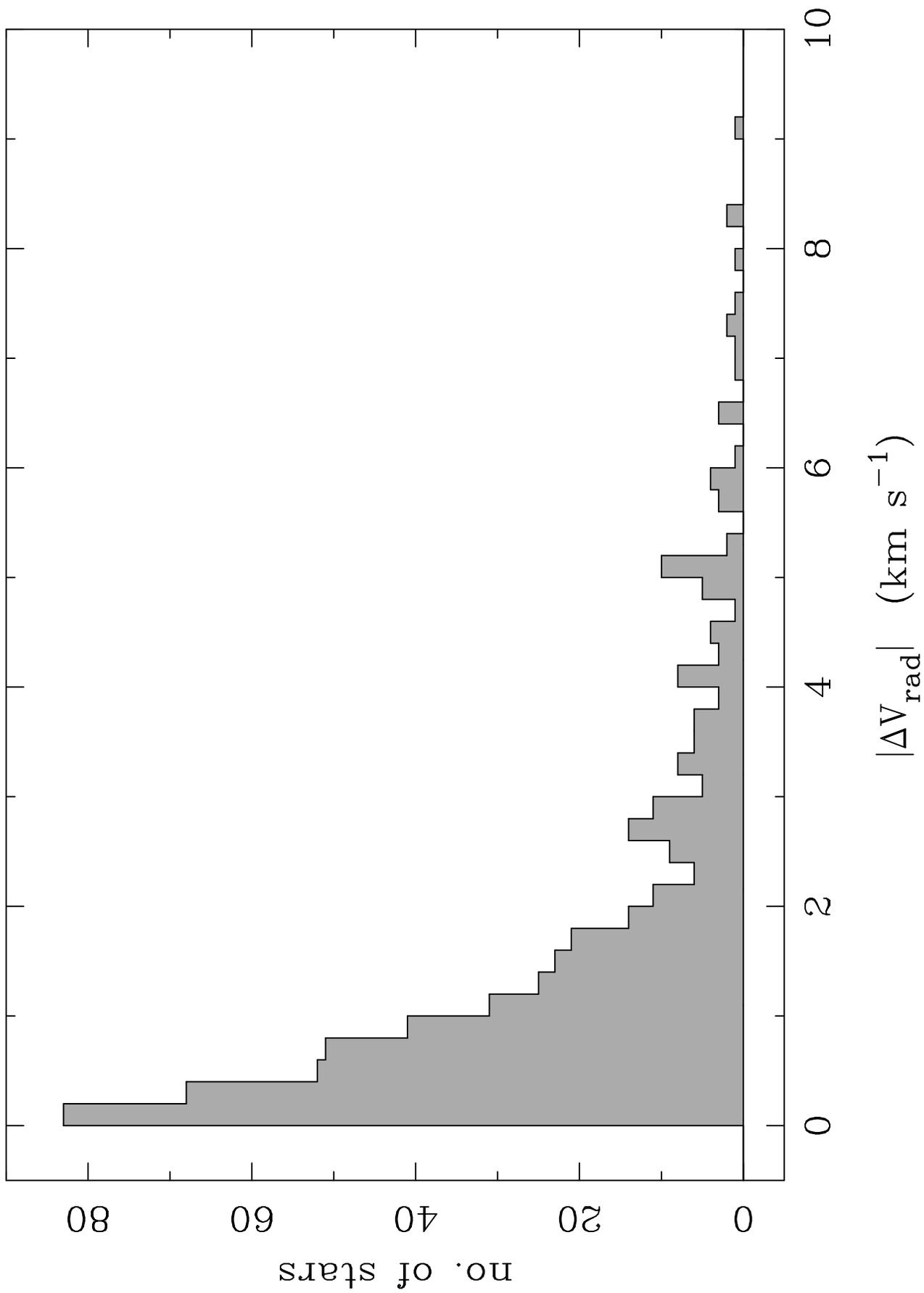
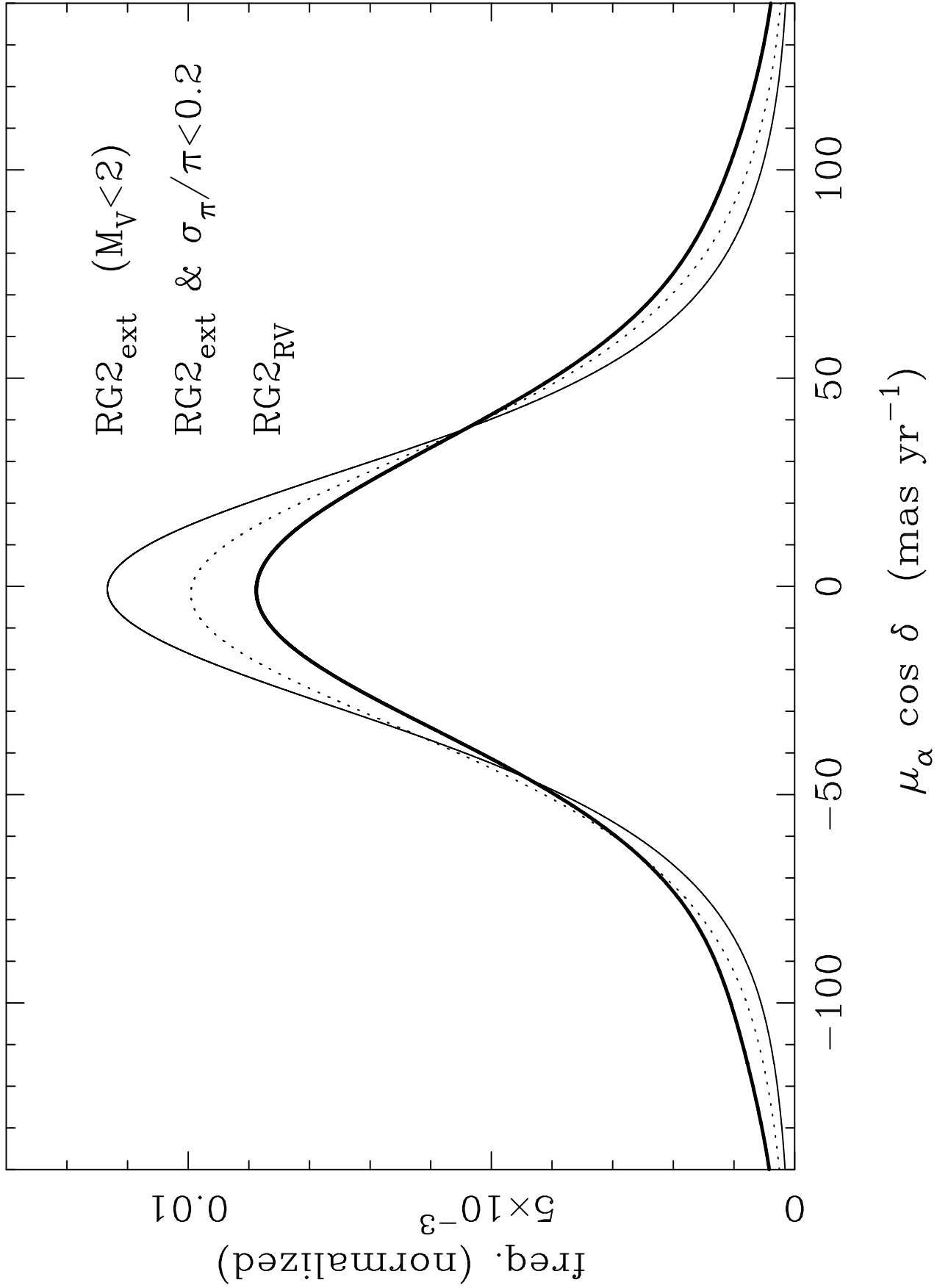
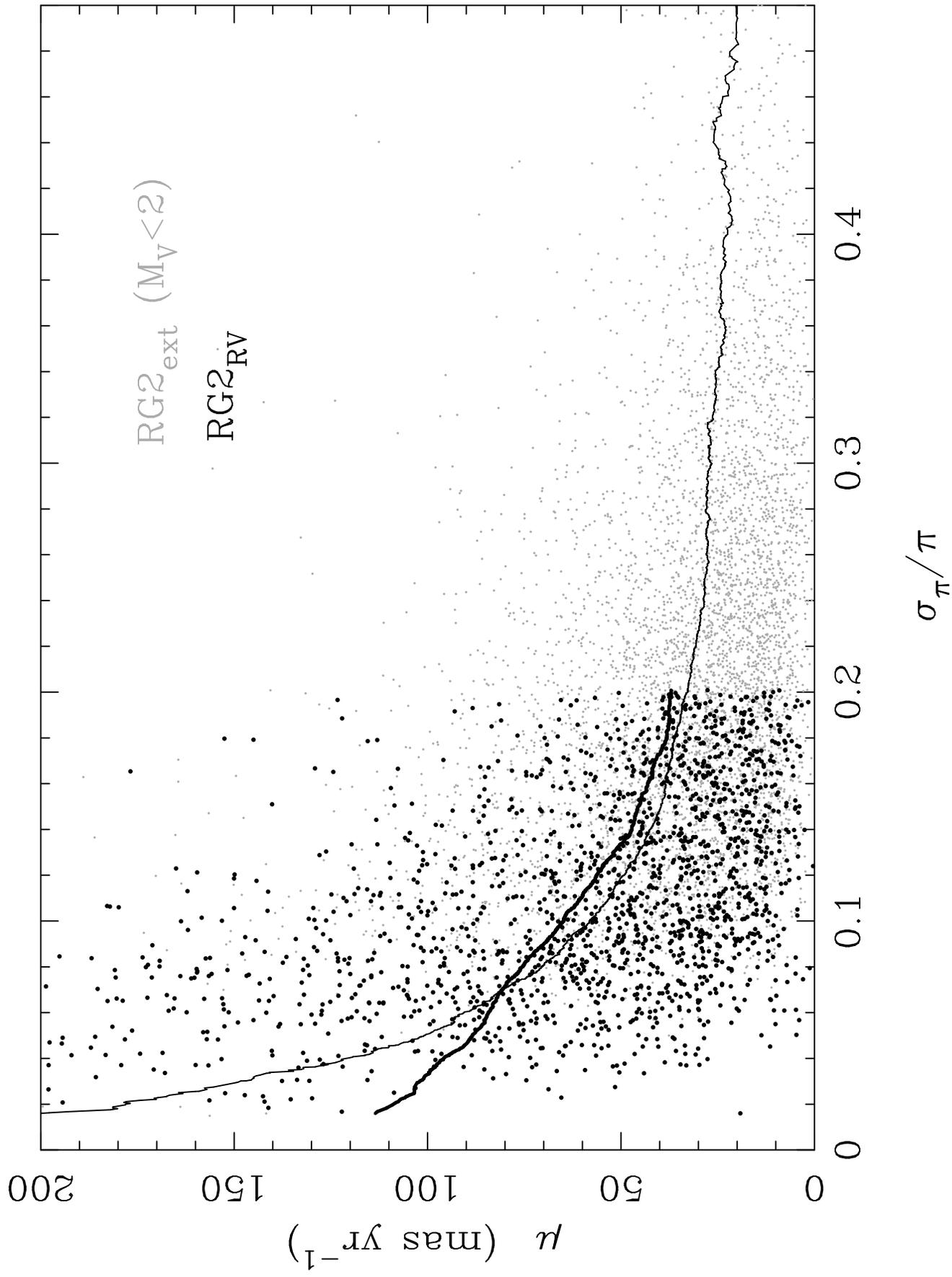
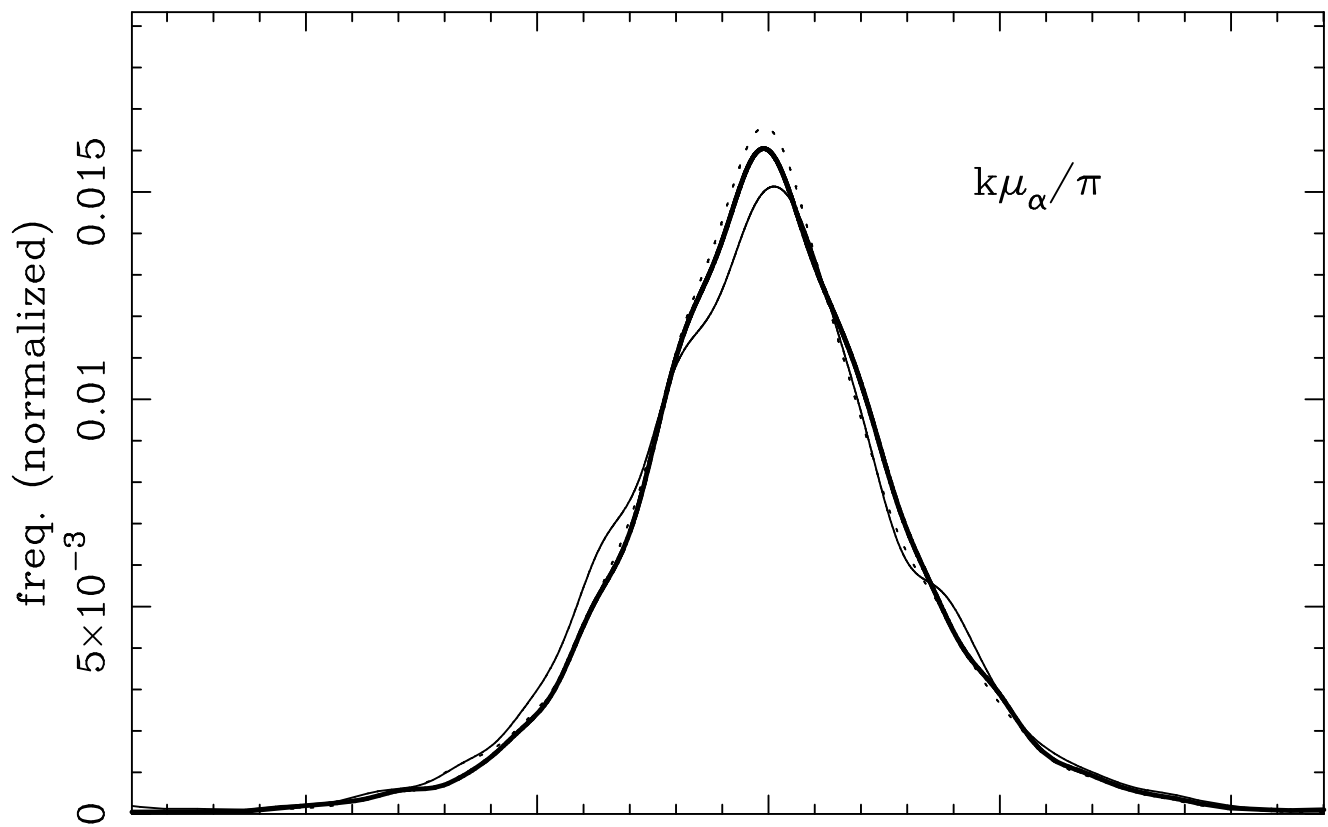
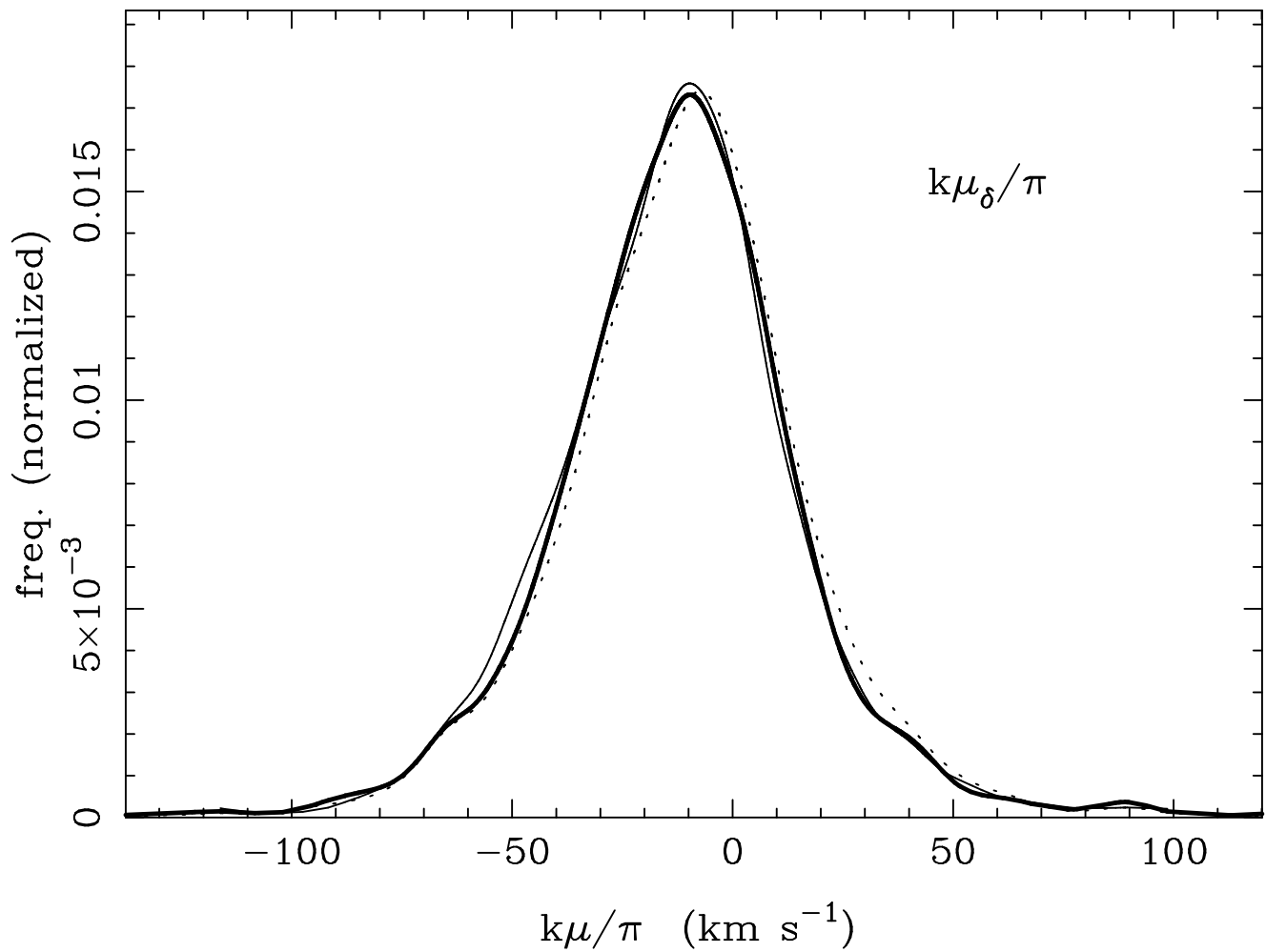
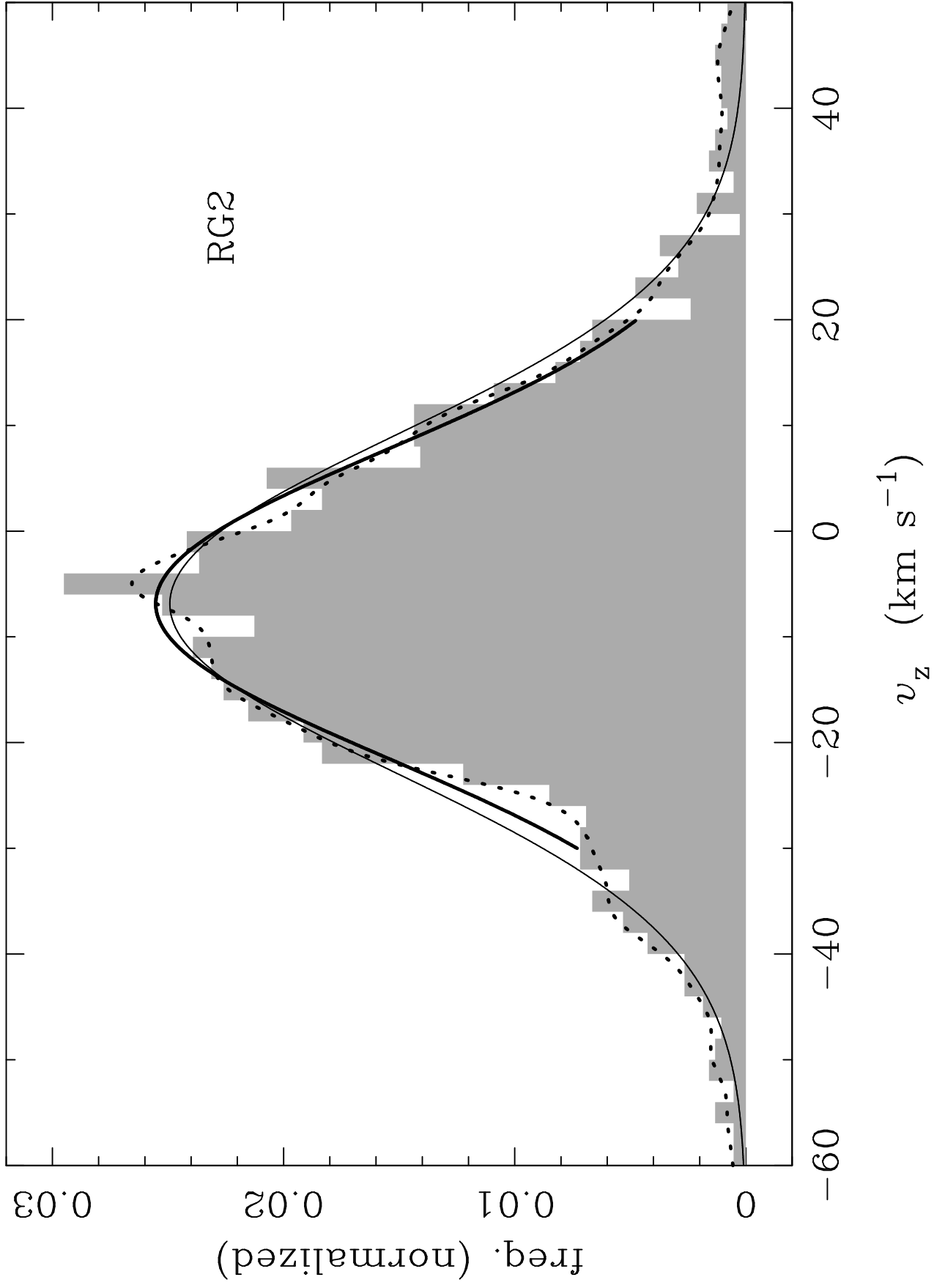
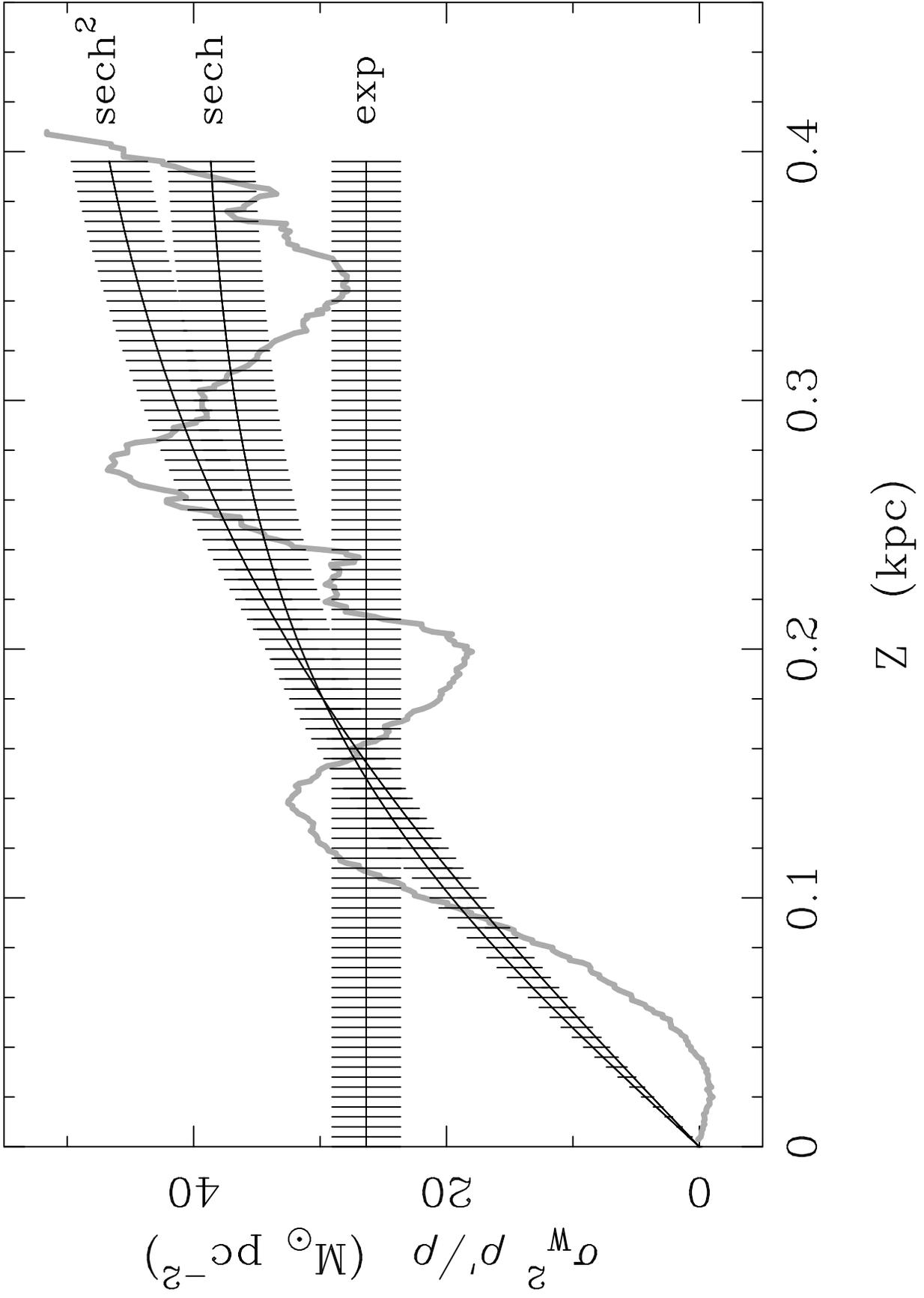
\end{document}